\documentclass[apj,numberedappendix,twocolappendix]{emulateapj}
\usepackage{apjfonts}

\usepackage{graphicx}
\usepackage{mathrsfs}
\usepackage{natbib}
\usepackage{amssymb}
\usepackage{bm}
\bibpunct[; ]{(}{)}{;}{a}{}{,}

\def\totd{{\mathrm{d}}}
\defcitealias{BS03}{BS03}

\shorttitle{RADIATION-DRIVEN MAGNETO-ACOUSTIC INSTABILITY (RMI)}
\shortauthors{FERN\'ANDEZ \& SOCRATES}

\begin{document}

\title{Nonlinear evolution of the radiation-driven magneto-acoustic instability (RMI)}
\author{Rodrigo Fern\'andez\altaffilmark{1} and Aristotle Socrates\altaffilmark{2}}
\affil{Institute for Advanced Study. Einstein Drive, Princeton, NJ 08540, USA.}
\altaffiltext{1}{Einstein Fellow}
\altaffiltext{2}{John N. Bahcall Fellow}

\begin{abstract}
We examine the nonlinear development of unstable magnetosonic waves driven by a
background radiative flux -- the Radiation-Driven Magneto-Acoustic Instability
(RMI, a.k.a. the ``photon bubble'' instability). The RMI may serve as a
persistent source of density, radiative flux, and magnetic field fluctuations
in stably-stratified, optically-thick media.  The conditions for instability
are present in a variety of astrophysical environments, and do not require the
radiation pressure to dominate or the magnetic field to be strong.  Here we
numerically study the saturation properties of the RMI, covering three orders of
magnitude in the relative strength of radiation, magnetic field, and gas
energies. Two-dimensional, time-dependent radiation-MHD simulations of local,
stably-stratified domains are conducted with Zeus-MP in the optically-thick,
highly-conducting limit. Our results confirm the theoretical expectations 
of Blaes and Socrates (2003) in that the RMI operates
even in gas pressure-dominated environments that are weakly magnetized.
The saturation amplitude is a monotonically increasing function of the ratio of radiation to 
gas pressure. Keeping this ratio constant, we find that the saturation
amplitude peaks when the magnetic pressure is comparable to the radiation pressure.
We discuss the implications of our results for the dynamics of magnetized stellar envelopes,
where the RMI should act as a source of sub-photospheric perturbations.
\end{abstract}


\maketitle

\section{Introduction}

Many gravitationally-bound astrophysical systems remove their binding energy
by the diffusion of radiation through optically thick regions.
Examples are main-sequence stars and accretion disks
onto compact objects. In general, these systems
tend to be good conductors and therefore support magnetic fields. 

\begin{figure}
\includegraphics*[width=\columnwidth]{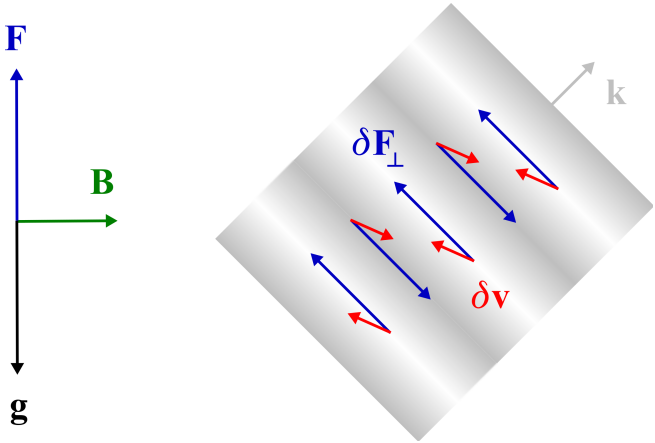}
\label{f:force_diagram}
\caption{Driving of magnetosonic modes by a background radiative flux $\mathbf F$ 
(\citetalias{BS03}). The shaded region represents a magnetosonic plane wave, with
darker and lighter regions corresponding to overdensities 
and rarefactions, respectively. The perturbation to the flux $\delta \mathbf{F}$ does work on the
component of the velocity perturbation $\delta \mathbf{v}$ perpendicular to the  
wave vector $\mathbf{k}$, provided that the latter is not entirely parallel 
or perpendicular to the magnetic field $\mathbf B$. The acceleration of gravity 
is denoted by $\mathbf g$, and velocity vectors are drawn for a slow mode in the
weak field limit.}
\end{figure}

\citet[hereafter BS03]{BS03} found that these environments
are susceptible to magnetosonic overstability.
The physical mechanism responsible for secular driving involves
short wavelength compressible fluctuations that grow exponentially due to the presence 
of a background radiative flux.  Fluid motion along the direction of the equilibrium magnetic field,
resulting from magnetic tension forces, couples to the radiative flux perturbation.  
In the event that these two perturbations are in phase, the radiation field performs work 
on the fluid oscillation and increases its amplitude (Figure~\ref{f:force_diagram}).  

A surprising result of \citetalias{BS03}
is that weakly magnetized and/or gas pressure dominated equilibria are subject
to the same radiative driving mechanism responsible for overstability in radiation pressure and/or magnetic 
pressure dominated environments.  The instability mechanism is so generic that even when the
radiation is degenerate (e.g., neutrinos in core-collapse supernova environments, \citealt{socrates2005}),
it may still operate.  Furthermore, a similar mechanism operates in the effect that the 
energy-transporting particles are charged and diffuse primarily along magnetic field
lines \citep{socrates2008}.     

Before \citetalias{BS03}, such phenomena were thought to be restricted to radiation-pressure-dominated
media that are strongly magnetized.  For these conditions, the instability is
often referred to as ``photon bubbles'' \citep{prendergast1973,arons1992,gammie1998,BS01,begelman2001},
though neither buoyancy, nucleation, nor surface tension play a role in the driving
mechanism.  
Numerical investigation of the nonlinear development of this instability has
been limited, with only a few calculations of 
strongly-magnetized and mostly radiation pressure dominated atmospheres
reported in the literature \citep{klein1996,hsu1997,davis2004,turner2004,turner2005,turner2007,jiang2012}.
A major difficulty for numerical studies results from the fact that the radiation diffusion time over 
a wavelength smaller than the gas pressure scale height needs to be resolved in order to 
attain numerical convergence \citep{turner2005}.  In practice,
the oscillation period itself is often much shorter than the e-folding time of the mode,
requiring a large number of time steps to capture the secular amplification of the waves.

Here we take advantage of the parallel radiation-MHD code 
Zeus-MP \citep{hayes2006} to begin an exploration of the more extended parameter 
space for unstable modes found by \citetalias{BS03}.
To reflect the physical character of the driving mechanism, we designate
it the \emph{Radiation-Driven Magneto-Acoustic Instability} (RMI).

The parameter space to be explored is motivated by the radiative
envelopes of massive stars, many of which are known
to support up to kG-strength magnetic fields (e.g., \citealt{wade2012} and references therein).
In main-sequence O stars, the radiation pressure only marginally exceeds the equilibrium gas
pressure; a magnetic field near equipartition with the gas 
provides favorable conditions for the growth of the RMI (e.g., \citetalias{BS03}, 
\citealt{turner2004}). Going down the main sequence in mass and 
luminosity, radiation pressure becomes increasingly less important, yet the instability may persist
near the photosphere.

The paper is organized as follows. Section 2 contains an overview of the
physics of the RMI. Section 3 describes our assumptions, the numerical
method employed, and the parameter space explored. Section 4 presents our results.
Section 5  briefly discusses implications for massive stars.
We conclude with a summary of our work in Section 6. 
Appendix~\ref{s:linear_derivation} contains a brief derivation of the growth
rate of the RMI and the stability criteria.
Appendix~\ref{s:code_tests} describes verification tests of Zeus-MP.
Mathematical symbols have their usual meanings; the most frequently used ones
are defined in Table~\ref{t:definitions}.

\section{Overview of the RMI}
\label{s: basics}

The photon bubble instability was discovered
numerically by \citet{arons1992} and \citet{gammie1998} in atmospheres that are radiation 
pressure dominated, with either strong ($B^2/8\pi\gg p_g$) 
or super-strong ($B^2/\pi\gg p_g+E/3$) magnetic fields.  
In their computations, they found the slow magnetosonic 
wave to be susceptible to over-stability, driven by the presence of a background 
radiative flux ${\bf F}$. \citet{BS01} subsequently found that
the fast magnetosonic mode could also be driven unstable, and that
in the presence of rotation, the splitting between the Alfv\'en and slow wave allows 
for the -- otherwise incompressible -- Alfv\'en branch to be driven as well. 
The effects of magnetic stratification in the magnetic- and radiation-dominated regime
have been addressed by \citet{tao2011}. 

The stability criteria and dynamics of the RMI for a constant magnetic field were derived
in \citetalias{BS03}. They found that (i) stability is determined by 
a balance between driving from the background flux ${\bf F}$, damping
due to radiative diffusion, and departures from thermal equilibrium, 
(ii) the instability can be thought of as magnetosonic waves that are secularly 
driven by the background flux ${\bf F}$, (iii) the instability mechanism operates
independently of the degree of thermal coupling between the gas and 
radiation, and (iv) the instability mechanism operates even when the magnetic and/or radiation
pressure is/are less than the equilibrium gas pressure.

In order to determine the relevant parameter space for numerical simulations of the RMI, 
we must understand the physics that determines whether a given atmosphere 
is unstable to RMI driving. A brief derivation
is found in Appendix~\ref{s:linear_derivation}. Below we provide a general overview. 

\begin{deluxetable}{cc}
\tabletypesize{\scriptsize}
\tablecaption{Frequently-Used Symbols\label{t:definitions}}
\tablewidth{0pt}
\tablehead{\colhead{Symbol} &\colhead{Quantity}}
\startdata
$\rho$      & density  \\
${\bf v}$   & velocity \\
$e$         & gas energy density\\
$E$         & radiation energy density\\
$T$         & temperature\\
$s$         & entropy per baryon\\
$n$         & baryon number density\\
$p_g$       & gas pressure\\
$c_i$       & isothermal gas sound speed\\
${\bf g}$   & gravitational acceleration\\
\noalign{\smallskip}
$p_{\rm r}$      & radiation pressure\\
${\bf F}$        & radiative flux\\
$\kappa_{\rm F}$ & flux-mean opacity\\
$\kappa_{\rm a}$ & absorption opacity\\
\noalign{\smallskip}
${\bf B}$        & magnetic field\\
${\bf v_A}$      & Alfv\'en velocity \\
$p_{\rm m}$      & magnetic pressure\\
\noalign{\smallskip}
${\bf k}$           & wave vector\\
$\omega$            & mode frequency \\
$\omega_{\rm diff}$ & diffusion frequency (eq.~[\ref{eq:omega_diff}])\\
$\omega_{\rm th}$   & heat exchange frequency (eq.~[\ref{eq:omega_thermal}])\\
${\tilde\omega}^2$ & $\omega^2-\left({\bf k}\cdot{\bf v_A}\right)^2$\\
\noalign{\smallskip}
$\Re$                 & rapid-diffusion parameter (eq.~[\ref{eq:rapid_diffusion_parameter}])\\
$\varepsilon_{\rm c}$ & critical Eddington ratio (eq.~[\ref{eq:critical_epsilon}])\\
$c$                   & speed of light\\
\noalign{\smallskip}
$L$                   & box size\\
$c_{\rm s0}$          & adiabatic gas sound speed at $z=0$\\
$\rho_0$              & density at $z=0$
\enddata
\end{deluxetable}

\vspace{0.5in}

\subsection{Characteristic Frequencies}
\label{ss: stability}

We consider here the evolution of plane waves of frequency $\omega$ and wave vector $\mathbf k$
on an ideal MHD fluid with constant background magnetic field $\mathbf B$ and gravitational
acceleration $\mathbf g$.  Furthermore, we take the various frequency-weighted opacities
to be constant in order to isolate the RMI from opacity-driven instabilities.

All of the unstable RMI parameter regimes found by \citetalias{BS03} involve 
heat rapidly diffusing across a wavelength in comparison
to the frequency of oscillation. That is
\begin{equation}
\label{eq:rapid_diffusion_condition}
\omega_{\rm diff}\gg \textrm{Re}(\omega),
\end{equation}
where 
\begin{equation}
\label{eq:omega_diff}
\omega_{\rm diff} = \frac{c\,k^2}{3\kappa_F\rho}
\end{equation}
is the radiative diffusion frequency, 
and $\kappa_F$ is the flux-mean opacity.  
Radiation diffuses rapidly in comparison to the mode oscillation 
time, i.e., the perturbations are highly non-adiabatic.

In this paper, we restrict ourselves to the case where
the gas and radiation are thermally well coupled, so that both are described 
by a single temperature. This condition is satisfied in the radiative envelopes of
massive stars. Quantitatively, we demand that
\begin{equation}
\label{eq:thermal_locking_condition}
\omega_{\rm th}\gg \textrm{Re}(\omega),
\end{equation}
where
\begin{equation}
\label{eq:omega_thermal}
\omega_{\rm th} = \frac{4(\gamma-1)E}{p_g}\,\kappa_a \rho c,
\end{equation}
is a characteristic heat exchange frequency \citepalias{BS03}, with $\kappa_a$ the thermal absorption 
(Planck-mean) opacity, and $c$ the speed of light.
In this limit, gas-acoustic perturbations travel at the isothermal
\footnote{This is not the same as using an isothermal equation of state, 
since $\omega_{\rm th}$ is a local quantity.} sound speed, $c_i^2 = p_g/\rho$.

\subsection{Stability Criteria}

As shown in Appendix~\ref{s:linear_derivation}, the criterion
for RMI driving is given by (eqns.~[\ref{eq:growth_rate_first_order}]-[\ref{e: easy_criteria}])
\begin{equation}
\label{eq:easy_criterion_opening}
k\,F\gtrsim \omega\,nT\left(-\frac{\partial s}{\partial \ln\rho}\right)_{\rm T}
             \,\frac{\left[\omega^2-(\mathbf{k}\cdot\mathbf{v}_{\rm A})^2\right]}{k^2v^2_A}.\nonumber
\end{equation}
The expression above quantifies the competition between 
the rate of radiative energy transfer into a fluctuation over its wavelength
$\lambda=2\pi/k$ due to the presence of a background flux $F$, and 
the rapid diffusion of energy that results from compression.  
Equation~(\ref{eq:easy_criterion_opening}) is obtained as a first order expansion in the small
parameters $(kH_{\rm g})^{-1}$, $\textrm{Im}(\omega)/\textrm{Re}(\omega)$, and $\textrm{Re}(\omega)/\omega_{\rm diff}$,
where $H_{\rm g}$ is the gas pressure scale height. 
Equation~(\ref{eq:easy_criterion_opening}) differs from equation~(\ref{eq:growth_rate_first_order}) only
by angular factors, being thus accurate to within a factor of order unity.
A more accurate value than equation~(\ref{eq:growth_rate_first_order}) can be obtained by
solving the eighth-order dispersion relation of \citetalias{BS03}.

For slow magnetosonic waves, we may write
\begin{equation}
\label{eq:slow_criterion}
F \gtrsim \zeta^3\left[p_{\rm g} + \frac{4}{3}E\right]\,c_i,
\end{equation}
where we have used equation~(\ref{eq:dsdrho}) and defined
\begin{equation}
\label{eq:zeta_mag}
\zeta \equiv \min\left(\frac{v_A}{c_i}, 1\right).
\end{equation}
The growth rate can be approximated by
\begin{equation}
\label{eq:growth_slow}
\Gamma_{\rm slow}  \simeq  \frac{\varepsilon g}{4c_i}\left( 1 + \frac{3p_{\rm g}}{4E}\right)
\zeta \left[1-\frac{\zeta^3}{F}\left(p_{\rm g}+\frac{4}{3}E\right)\,c_i\right],
\end{equation}
where
\begin{equation}
\label{eq:eddington_ratio}
\varepsilon\equiv\frac{\kappa_F F}{cg}\equiv \frac{F}{F_{\rm edd}}
\end{equation}
is the Eddington ratio. The angular dependence of equation~(\ref{eq:growth_rate_first_order}) 
has been replaced by a global factor of $1/2$ that accounts for the
product of the sines of the relative angles between $\mathbf{k}$, $\mathbf{B}$,
and $\mathbf{F}$. 

For fast magnetosonic waves, the corresponding instability criterion reads
\begin{equation}
\label{eq:fast_criterion}
F \gtrsim \frac{1}{\zeta^3}\left[p_{\rm g} + \frac{4}{3}E\right]\,v_A
\end{equation}
with a corresponding asymptotic growth rate
\begin{equation}
\label{eq:growth_fast}
\Gamma_{\rm fast} \simeq \frac{\varepsilon g}{4c_i}\left( 1 + \frac{3p_{\rm g}}{4E}\right)
\left(\frac{c_i}{v_A} \right)\zeta^3
\left[1-\frac{\zeta^{-3}}{F}\left(p_{\rm g}+\frac{4}{3}E\right)\,v_A\right].
\end{equation}
The extra factor of $1/2$ has also been included.

\begin{figure}
\includegraphics*[width=\columnwidth]{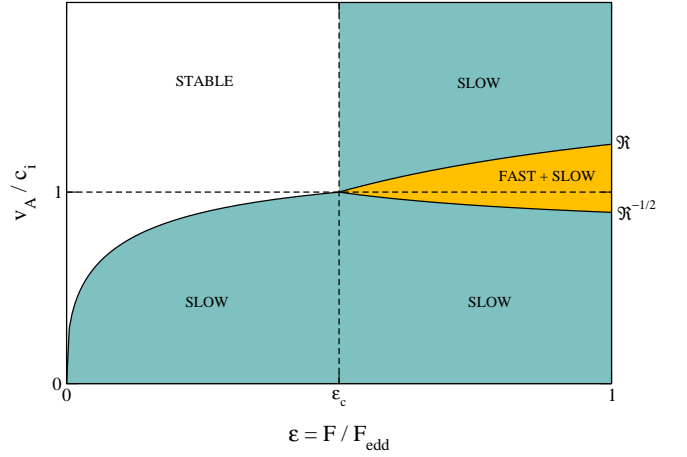}
\label{f:parameter_diagram}
\caption{Unstable parameter regions in the thermally locked case (mode
frequencies smaller than eq.~[\ref{eq:omega_thermal}]).  If the
``rapid-diffusion'' parameter $\Re$ (eq.~[\ref{eq:rapid_diffusion_parameter}])
exceeds unity, then a critical value of the Eddington ratio $\varepsilon_c$
(eq.~[\ref{eq:critical_epsilon}]) exists. For $\varepsilon > \varepsilon_c$,
unstable fast-magnetosonic modes are possible, and slow-magnetosonic modes are
unstable for any magnetic field. Below $\varepsilon_c$, unstable slow modes are
possible only up to a maximum value of $v_A/c_i$ given by
equation~(\ref{eq:slow_criterion}), and fast modes are stable.  
If $\Re < 1$, then for any Eddington ratio
there exists a maximum value of $v_A/c_i < 1$ for which slow modes
can be destabilized.}
\end{figure}

It is instructive to visualize the instability criteria for slow and
fast waves
in a plane with the Eddington ratio $\varepsilon$ on the x-axis and the ratio $v_A/c_i$
on the y-axis, as is shown in Figure~\ref{f:parameter_diagram}.
The critical stability curves can be obtained by relating the Eddington ratio to the
ratio of radiation to gas pressure through
\begin{equation}
\label{eq:varepsilon_pratio}
\ell_H\, \frac{E/3}{p} = \frac{\varepsilon}{1-\varepsilon}.
\end{equation}
where
\begin{equation}
\ell_H = \frac{H_g}{H_r}
\end{equation}
is the ratio of gas pressure scale height to radiation pressure scale height $H_r$.

The resulting curves have an additional free parameter,
\begin{equation}
\label{eq:rapid_diffusion_parameter}
\Re = \frac{\ell_H}{4}\,\frac{c/c_i}{\tau_p},
\end{equation}
where $\tau_p = \kappa\rho H_g$ is the optical depth over a gas pressure
scale height. The parameter $\Re$ is proportional to the ratio of diffusion speed $c/\tau_p$
to the isothermal sound speed. When $\Re > 1$, there
is a critical value of the Eddington ratio
\begin{equation}
\label{eq:critical_epsilon}
\varepsilon_c = \left[1 + \frac{c/c_i}{\tau_p} - \frac{4}{\ell_H}\right]^{-1}
\end{equation}
above which fast and slow waves are unstable for any magnetic field strength. If $\Re < 1$, then
only slow modes can be destabilized, and for every Eddington ratio there
is a maximum field strength for which the slow wave RMI operates.

\subsection{Aspects of the Stability Criteria}
\label{s:aspects}

The growth/damping rate and stability criteria above give insight 
regarding the properties of the equilibrium about which the RMI
takes place.  Balance between radiative driving and damping is, 
respectively, a balance between energy input from the background radiative
flux ${\bf F}$ and energy loss due to diffusive cooling that results from compression, 
which is $\propto(\partial s/\partial\rho)_T$.  

Consider, for example, the slow magnetosonic wave in the strong field limit,
so that equation (\ref{eq:easy_criterion_opening}) becomes
\begin{eqnarray}
\frac{1}{\lambda_{k}}F & \gtrsim  & 
         \frac{1}{\Delta t_{\omega}}\,n T \left(-\frac{\partial s}{\partial \ln\rho}\right)_T\nonumber\\ 
\frac{1}{\lambda_k} \frac{Ec}{\tau_p} & \gtrsim &
     \frac{1}{\Delta t_\omega}\left(p_{\rm g} + 4p_{\rm r}\right)\nonumber\\
\frac{E}{\Delta t_{\rm diff}} & \gtrsim & \frac{(p_{\rm g}+4p_{\rm r})}{\Delta t_\omega}
\end{eqnarray}
where $\lambda_k=2\pi/k$ is the wavelength of an oscillation with wave number $k$,
and $\Delta t_{\omega}=2\pi/\omega$ is the oscillation period.  
The left hand side 
is the upper limit -- if, e.g., the perturbations were nonlinear -- for the rate of radiative energy 
transfer into a fluctuation of scale $\lambda_k$.  The characteristic velocity at which the background 
radiative field transfers radiative energy from one layer to another is given by $c/\tau_{\rm p}$,
and the corresponding time at which radiative diffusion transfers energy across the 
wavelength $\lambda_k$ is given by $\Delta t_{\rm diff}$.  The right hand side of the criteria 
for instability results from entropy or pressure perturbations that are sourced by changes
in density.  Since radiation diffuses, effectively, instantaneously over a wavelength $\lambda_k$
during the oscillation time $\Delta t_{\omega}$, the rate of energy loss per unit volume is 
capped by the relative contribution to pressure or entropy changes that arise from 
compression, which is $\propto (\partial s/\partial \rho)_T$ and the oscillation time 
$\Delta t_{\omega}$.

By re-inserting $\Xi$ (eq.~[\ref{eq:Xi_app_definition}]), the instability criteria can
be written as
\begin{eqnarray}
\frac{\Delta t_{\omega}}{\Delta t_{\rm diff}} & \gtrsim & \frac{1}{\varepsilon}
   \,\Xi\left[\omega,{\bf k}, {\bf B}\right]\nonumber\\
\label{eq:instability_timescales}
 \frac{c/\tau_p}{v_{\rm ph}} & \gtrsim&\frac{1}{\varepsilon}\,\Xi\left[\omega,{\bf k}, {\bf B}\right]
\end{eqnarray}
where the phase velocity $v_{\rm ph}=\omega/k$ (eq.~[\ref{eq:disprel_magnetosonic}]), 
and we have taken $\varepsilon p_{\rm g}/p_{\rm r} \sim 1$.
The maximum value of the left hand side is achieved at the photosphere where $\tau=1$.  

Equation~(\ref{eq:instability_timescales}) re-enforces several facts that are important in 
attempting to understand where RMI phenomena take place.  Instability vanishes in limit 
$\tau_p\rightarrow\infty$ since the ratio of the radiation flux to energy density vanishes as well.  
The free energy for the instability is rooted in the anisotropy of the radiation field, which 
leads to a radiative flux that can then perform work on fluid fluctuations.  Therefore, in 
the center or stars (or, e.g., at the surface of last scattering for the cosmic background radiation) 
RMI driving  does not operate. Consequently, instability favors regions relatively near the photosphere.

The presence of $v_{\rm ph}$ in the denominator of the left hand side implies that the instability favors
magneto-acoustic oscillations with small phase velocity.  Given the oscillation wavelength, the rate
at which radiation removes energy from a fluid disturbance is given by 
the phase velocity.  That is, the rate at which radiative diffusion can remove energy 
from the wave is relatively small for relatively slow oscillations.

\section{Numerical Setup}

Our aim is to follow the nonlinear development of the RMI in a small
patch of stellar interior.
Below we describe the equations solved, numerical
method, initial conditions, boundary conditions, and choice of
parameters for simulations.  

\subsection{Equations}

We solve the radiation-magnetohydrodynamic equations in the ideal MHD and grey diffusion limits
\begin{eqnarray}
\label{eq:mass_conservation}
\frac{\partial \rho}{\partial t} &+& \nabla \cdot \left(\rho\mathbf{v}\right)  = 0 \\
\label{eq:momentum_conservation}
\frac{\partial \mathbf{v}}{\partial t} + \mathbf{v}\cdot \nabla \mathbf{v}  & =  & -\frac{1}{\rho}\nabla p_{\rm g}
+ \frac{1}{4\pi\rho}\left(\nabla \times \mathbf{B} \right)\times \mathbf{B} + \frac{\kappa_F}{c}\mathbf{F} + \mathbf{g}\\
\label{eq:fluid_energy_conservation}
\frac{\partial e}{\partial t} + \mathbf{v}\cdot\nabla e & = & -\gamma e\nabla\cdot\mathbf{v}+ 
                                \kappa_{\rm a} \rho c\, \left(E - a T_{\rm g}^4\right)\\
\label{eq:radiation_energy_conservation}
\frac{\partial E}{\partial t} + \mathbf{v}\cdot \nabla E & = & -\frac{4}{3}E\nabla\cdot\mathbf{v}
                   -\kappa_{\rm a} \rho c\, \left(E- a T_{\rm g}^4\right) -\nabla\cdot \mathbf{F} \\
\label{eq:radiation_diffusion}
\mathbf{F} & = & -\frac{1}{3}\frac{c}{\kappa_{\rm F}\rho}\, \nabla E\\
\label{eq:induction_equation}
\frac{\partial \mathbf{B}}{\partial t}  & = & \nabla \times \left( \mathbf{v}\times \mathbf{B}\right).
\end{eqnarray}

Our goal is to isolate the growth and nonlinear development of 
the RMI, which serves as justification for setting the flux-mean opacity
$\kappa_{\rm F}$ to a constant, as in the case of the Thomson scattering opacity.
Radiation-hydrodynamic instabilities such as the $\kappa -$mechanism
and the so-called strange mode instability (e.g., \citealt{glatzel1994}), which rely on opacity 
variations, are thus excluded.

The Planck-mean and intensity-mean opacities normally
enter independently into the radiation and gas energy equations 
(eqns.~[\ref{eq:fluid_energy_conservation}]-[\ref{eq:radiation_energy_conservation}]).
Here they are set equal to one another, implicitly assuming that the radiation
follows a Planck distribution with its own temperature (e.g., \citetalias{BS03}).  
We refer to these opacities collectively
as the absorption opacity $\kappa_a$. In our calculations, the ratio $\kappa_a/\kappa_F$ 
is fixed to a value such that the condition for thermal locking (eq.~[\ref{eq:thermal_locking_condition}]) 
is satisfied.

Equations (\ref{eq:mass_conservation})-(\ref{eq:induction_equation}) are closed by 
an equation of state for the matter fluid, which we take to be an ideal gas of adiabatic index $\gamma$.
It follows that $p_{\rm g} = \rho kT_g / (\mu m_p) = (\gamma-1) e$, where $\mu$ and $m_p$ are the mean molecular
weight and proton mass, respectively. 
In the diffusion approximation, the radiation stress tensor $\mathbb{P}$ is related to the radiation energy density
by $\mathbb{P}/E = \mathbb{I}/3$, where $\mathbb{I}$ is the identity tensor. We denote
the scalar radiation pressure by $p_{\rm r} = E/3$.
No flux-limiters are employed (eq.~[\ref{eq:radiation_diffusion}]).

A cartesian coordinate system is adopted, with the acceleration of gravity pointing in the negative $z$ direction.
Throughout this study, we restrict ourselves to two-dimensional simulations, with $x$ denoting the
coordinate direction transverse to gravity.

\subsection{Numerical Method}

To evolve the system of equations~(\ref{eq:mass_conservation})-(\ref{eq:induction_equation}), we employ the 
publicly available finite-difference code Zeus-MP \citep{hayes2006}. 
The MHD part is evolved using the
Modified Characteristics -- Constrained Transport method of \citet{hawley1995}. Shocks are captured
with the artificial viscosity prescription of \citet{vonneumann50}.  The radiation part is 
evolved implicitly via the Conjugate Gradient method \citep{hayes2006}, and interacts with the MHD sector
via operator splitting.

As a basic test of our implementation, we verify that the initial conditions (\S\ref{s:initial_conditions}) 
do not evolve when unperturbed. We have also tested the elimination of the flux limiter in
Zeus-MP by solving the diffusion equation in a unit square with periodic boundary conditions
and constant initial $E$. 
A more stringent test involves comparing eigenfrequencies of
magnetosonic eigenmodes on a uniform background in the thermally-locked limit 
with the asymptotic values found by \citetalias{BS03}. Details on both of these tests are provided 
in Appendix~\ref{s:code_tests}.

\begin{deluxetable*}{cccccccccccc}
\tablecaption{Models Evolved and Time-Averaged Properties\label{t:results}}
\tablewidth{\textwidth}
\tablehead{
\colhead{$p_{\rm r}/p_{\rm g}$\tablenotemark{a}}&
\colhead{$p_{\rm m}/p_{\rm g}$} &
\colhead{$\Re$\tablenotemark{b}} &
\colhead{$\varepsilon/\varepsilon_c$\tablenotemark{b}} &
\colhead{$L_x/H_p$\tablenotemark{c}} &
\colhead{$\theta_{BF}$\tablenotemark{d}} &
\colhead{$n_x\times n_z$\tablenotemark{e}} &
\colhead{$\bar{E}_{\rm k,tot}/(\bar{E}_{\rm r}+\bar{E}_{\rm g})$} &
\colhead{$\bar{E}_{\rm k,v}/\bar{E}_{\rm k,h}$} &
\colhead{$\log_{10}(\sigma_\rho / \bar\rho)$\tablenotemark{f}} &
\colhead{$\log_{10}(\sigma_{\rm F}/ \bar{F})$} &
\colhead{$\log_{10}(\sigma_{\rm B} / \bar{B})$}
}
\startdata
0.01  & 0.3  & 31 & 1.2 & 2 & $90^\circ$ & $128^2$ & $(2\pm 1.0)10^{-5}$ & $10^{-0.6}$ & $-3.0/-2.6/-1.7$ 
	                		 & $-3.0/-2.9/-2.0$ & $-2.6/-1.9/-1.9$ \\  
0.03  &      &    & 3.5 &   &            &         & $(3\pm 2.0)10^{-5}$ & $10^{-0.5}$ & $-2.9/-2.5/-1.6$ 
	              			 & $-2.8/-2.9/-1.9$ & $-2.5/-1.8/-1.8$ \\  
0.1   &      &    & 11  &   &            &         & $(7\pm 3.0)10^{-5}$ & $10^{-0.4}$ & $-2.7/-2.3/-1.5$ 
	            			 & $-2.6/-2.7/-1.8$ & $-2.2/-1.6/-1.7$ \\  
0.3   &      &    & 28  &   &            &         & $(2\pm 1.0)10^{-4}$ & $10^{-0.2}$ & $-2.1/-1.7/-1.3$ 
	              			 & $-2.2/-2.0/-1.5$ & $-1.5/-1.3/-1.5$ \\  
1     &      &    & 60  &   &            &         & $(3\pm 2.0)10^{-3}$ & $10^{-0.4}$ & $-1.3/-1.0/-0.8$ 
	              			 & $-1.7/-1.4/-1.0$ & $-0.7/-0.6/-0.9$ \\  
3     &      &    & 91  &   &            &         & $(1\pm 0.6)10^{-2}$ & $10^{-0.5}$ & $-0.7/-0.6/-0.5$ 
                                         & $-1.0/-1.0/-0.9$ & $+0.3/-0.1/-0.4$ \\  
10~\tablenotemark{g} & & & 110 & & &               & $>10^{-1.7}$ & \nodata & \nodata & \nodata & \nodata\\ 
\noalign{\medskip}
1     & 0.01 & 5.7 & 9.8 & 2 & $90^\circ$ & $128^2$ & $(5\pm 3.0)10^{-10}$ & $10^{-0.2}$ & $-4.6/-4.9/-4.2$ 
		              	    	      & $-4.4/-4.4/-4.4$ & $-4.5/-4.3/-3.5$ \\  
      & 0.1  & 18  & 34  &   &            &         & $(1\pm 0.5)10^{-3}$ & $10^{-0.4}$ & $-1.7/-1.4/-1.2$ 
		              		      & $-1.9/-1.7/-1.4$ & $-0.6/-0.6/-0.7$ \\  
      & 1    & 40  & 79  &   &            &         & $(5\pm 2.0)10^{-3}$ & $10^{-0.8}$ & $-1.0/-0.8/-0.6$ 
		              		      & $-1.3/-1.0/-0.9$ & $-0.9/-1.0/-1.2$ \\  
      & 10   & 40  & 79  &   &            &         & $(4\pm 0.4)10^{-3}$ & $10^{-3.0}$ & $-0.8/-0.7/-0.7$ 
					      & $-1.4/-1.3/-1.5$ & $-1.8/-2.0/-2.4$ \\  
\noalign{\medskip}
0.1   & 0.3  & 31 & 11 & 2 & $90^\circ$ & $16^2$  & $(2\pm 0.7)10^{-6}$ & $10^{-0.6}$ & $-3.0/-2.7/-2.1$ 
						  & $-3.1/-3.0/-2.5$ & $-2.6/-2.3/-2.3$\\
      &      &    &    &   &            & $32^2$  & $(4\pm 2.0)10^{-5}$ & $10^{-0.4}$ & $-2.7/-2.4/-1.5$ 
						  & $-2.8/-3.1/-1.9$ & $-2.3/-1.8/-1.8$\\
      &      &    &    &   &            & $64^2$  & $(6\pm 3.0)10^{-5}$ & $10^{-0.4}$ & $-2.7/-2.4/-1.5$ 
						  & $-2.7/-2.8/-1.8$ & $-2.2/-1.7/-1.7$\\
\noalign{\medskip}
0.1   & 0.3  & 31 & 11 & 4 & $90^\circ$ & $128\times 64$ & $(1.5\pm 0.5)10^{-4}$ & $10^{-0.6}$ & $-2.3/-2.1/-1.4$ 
					                 & $-2.6/-2.4/-1.7$ & $-1.9/-1.5/-1.5$ \\
      &      &    &    & 8 &            & $256\times 64$ & $(1.0\pm 0.5)10^{-4}$  & $10^{-0.5}$  & $-2.5/-2.2/-1.4$  
							 & $-2.6/-2.6/-1.7$ & $-2.1/-1.6/-1.6$\\
\noalign{\medskip}
1     & 0.3  & 31 & 60 & 2 & $75^\circ$ & $128^2$ & $(7\pm 3.0)10^{-4}$ & $10^{-0.9}$ & $-1.5/-1.3/-1.0$ 
					  & $-1.5/-1.3/-1.3$ & $-1.2/-1.1/-1.2$ \\  
      &      &     &     &   & $60^\circ$ &         & $(2\pm 1.0)10^{-3}$ & $10^{-0.8}$ & $-1.1/-1.0/-0.8$ 
					  & $-1.2/-1.2/-1.1$ & $-0.9/-0.7/-0.8$ \\  
\enddata
\tablenotetext{a}{Values of $p_{\rm r}/p_{\rm g}$, $p_{\rm m}/p_{\rm g}$, $\Re$, and
                  $\varepsilon/\varepsilon_{\rm c}$ at the center of the box.}
\tablenotetext{b}{The parameters $\Re$ and $\varepsilon_{\rm c}$ are determined by
		  requiring that $\omega_{\rm diff}$ is 100 times the slow magnetosonic
		  frequency of a mode with wavelength equal to $H_{\rm g}$, at domain center.}
\tablenotetext{c}{Width of the computational box in the direction transverse to gravity,
		  in units of $H_{\rm g}$ at $z=0$. The vertical size is always $2H_{\rm g}$.}
\tablenotetext{d}{Angle between the imposed magnetic field and the background radiative flux.}
\tablenotetext{e}{Resolution in the directions transverse and parallel to gravity, respectively.}
\tablenotetext{f}{Ratio of the root-mean-square fluctuation over the mean 
(eqns.~[\ref{eq:mean_value}]-[\ref{eq:rms_fluctuation}]) at positions $z/L_z=\{0.2,0.5,0.8\}$, respectively.}
\tablenotetext{g}{The diffusion solver breaks down before saturation, due to order unity density fluctuations.}
\end{deluxetable*}

\subsection{Equilibrium Structure and Initial Conditions}
\label{s:initial_conditions}

We initialize the computational domain with an atmosphere in hydrostatic, radiative, and thermal equilibrium.
The equations describing this atmosphere are obtained from equations~(\ref{eq:momentum_conservation})-(\ref{eq:radiation_diffusion}) 
and the gas equation of state, by setting $\mathbf{v} = 0$, $aT_{\rm g}^4 = E$, and dropping the time derivatives:
\begin{eqnarray}
\label{eq:momentum_hydrostatic}
\frac{\totd p_{\rm g}}{\totd z}       & = & -(1-\varepsilon) \rho g\\
\label{eq:radiation_hydrostatic}
\frac{\totd E}{\totd z}       & = & -3\varepsilon \rho g\\
\label{eq:eos_hydrostatic}
\frac{\totd \ln\rho}{\totd z} & = & \frac{\totd \ln p_{\rm g}}{\totd z} - \frac{1}{4}\frac{\totd \ln E}{\totd z}.
\end{eqnarray}
The equilibrium magnetic field ${\bf B}$ is taken to be uniform in space and therefore 
does not contribute to the equilibrium structure of the atmosphere.

The problem is completely specified by nine parameters at one location. These are 
the ratio of box size $L$ to the gas pressure scale height $H_g$; the Eddington ratio $\varepsilon$;
the ratio of radiation to gas pressure $p_{\rm r}/p_{\rm g}$; the adiabatic index $\gamma$; the mean molecular
weight $\mu$, the ratio of isothermal sound speed to light speed $c_i/c$; the optical depth over a gas
pressure scale height $\tau_p$; the ratio of magnetic to gas pressure 
$p_{\rm m}/p_{\rm g} = B^2/(8\pi p_{\rm g})$; and the angle between magnetic field and 
radiative flux $\cos\theta_{\rm BF} = \mathbf{B\cdot F}/(|\mathbf{B}||\mathbf{F}|)$. 

We initialize the domain in such a way that there is cancellation to within a few parts in 
$10^{-6}$ between all the forces. This is possible because Zeus-MP computes gradients using finite
differences on a staggered mesh. Evolving this initial state without additional 
perturbations excites vertical fast magnetosonic modes, which decay
(there is no RMI driving if the wave-vector is perpendicular to the magnetic field and/or
parallel to the radiative flux).

Random initial perturbations in density are applied from $z=0.1L_z$ to $z = 0.9L_z$ for
all $x$. The amplitude is $10^{-6}$, and the random number is
varied from cell to cell. The internal energy density is also perturbed following the
density, $\delta e/e = \delta \rho/\rho$, so that the gas temperature and radiation
energy density remain uniform everywhere. 

For simplicity, we consider atmospheres with $\ell_H = 1$, which are stably-stratified 
in the rapid diffusion limit \citepalias{BS03}.

\subsection{Boundary Conditions}

At the inner vertical boundary, we fix the density, gas-, and radiation energy density to their
steady-state values, by continuing the initial solution beyond the boundary.
The radial velocity is reflected at this boundary, while the tangential
velocity is set to zero in the ghost cells. The magnetic field is set to its
uniform initial value.  This arrangement ensures that all waves are reflected
at the inner $z$ boundary, and that the domain does not collapse under the
action of gravity. A constant radiative flux $F$ is thus incident on the box
from below.

At the outer $z$ boundary, we use an outflow boundary condition that allows
waves to leave the domain but avoids runaway mass loss. To account for the fact
that this is an internal patch inside a hydrostatic structure rather than a
domain with a `vacuum' external boundary, we set the outflow radial velocity so
that the mass flux leaving the domain is that due do the excess density over
the steady-state value. At a given transverse coordinate, we thus set
\begin{equation} v_{r,\rm ghost} = \max\left(v_{r,a},0\right) \frac{\max\left(
\rho_a - \rho_0, 0\right)}{\rho_a}, \end{equation} where the subscripts $0$ and
$a$ label values in the initial and time-dependent solution at the last active
cell inside the outer radial boundary.  This prescription automatically
includes the case where the density is much larger than the steady-state value,
smoothly approaching a pure outflow boundary condition for the radial velocity.
The density, gas energy density, magnetic field, and tangential velocity from
the last active zone are copied into the two ghost zones. If the radiative flux
between the last pair of active zones is positive, the radiation energy density
in the ghost zones is set so as to keep this flux constant across cells. If
instead the flux is zero or negative, then the radiation energy density is set
to have zero gradient, yielding zero flux. 

The horizontal boundary condition is periodic for all variables. 

\subsection{Models Evolved}
\label{s:model_parameters}

We run two base sequences of models that vary the ratios of radiation to gas
pressure $p_{\rm r}/p_{\rm g}$ and magnetic to gas pressure $p_{\rm m}/p_{\rm
g}$.  These base sequences employ square computational boxes of two gas
pressure scaleheights on a side $(L_x=L_z=2H_p)$, with a resolution of
$128\times 128$ cells, and a time step $\Delta t$ equal to the diffusion time
for a wavelength $H_{\rm g}/8$.  The magnetic field points in the direction
perpendicular to gravity $(\hat x)$.  This choice of field geometry allows waves
to cross the domain many times while their amplitude grows. We perform
a few runs that explore the effects of changing the box size, resolution, and
field orientation. All models are shown in Table~\ref{t:results}, with
parameters showing values at the \emph{center of the domain}.

In all simulations, we take $c/c_i = 1000$, $\gamma=5/3$, and $\mu=1$.  The
optical depth in the box is chosen so that the diffusion frequency $\omega_{\rm
diff}$ is $100$ times larger than the slow magnetosonic frequency for a
wavelength equal to $H_{\rm g}$. The wave number dependence of $\omega_{\rm
diff}$ ensures that all modes of shorter wavelength are well within the
rapid-diffusion limit.  This choice fixes the value of $\Re$ and thus
$\varepsilon/\varepsilon_c$.  

The ratio $\kappa_{\rm a}/\kappa_{\rm F}$ is chosen so that $\omega_{\rm th}$
is $1000$ times larger than the slow magnetosonic mode of wavelength equal to
$H_{\rm g}/16$. Our choice is motivated by results from tests of radiation-MHD
waves on a uniform background (Appendix~\ref{s:code_tests}).


\section{Results}

\subsection{Linear Growth and Saturation}
\label{s:evolution}

The general behavior of all models involves
damping of the initially applied perturbations, and the subsequent 
emergence of unstable propagating waves. 
Because the background magnetic field is perpendicular to gravity
in most models, waves are allowed to cross the domain many times\footnote{To lowest order, the
group velocity of slow magnetosonic modes lies along the magnetic field.}.
This leads to exponential growth of the kinetic energy.

\begin{figure}
\includegraphics*[width=\columnwidth]{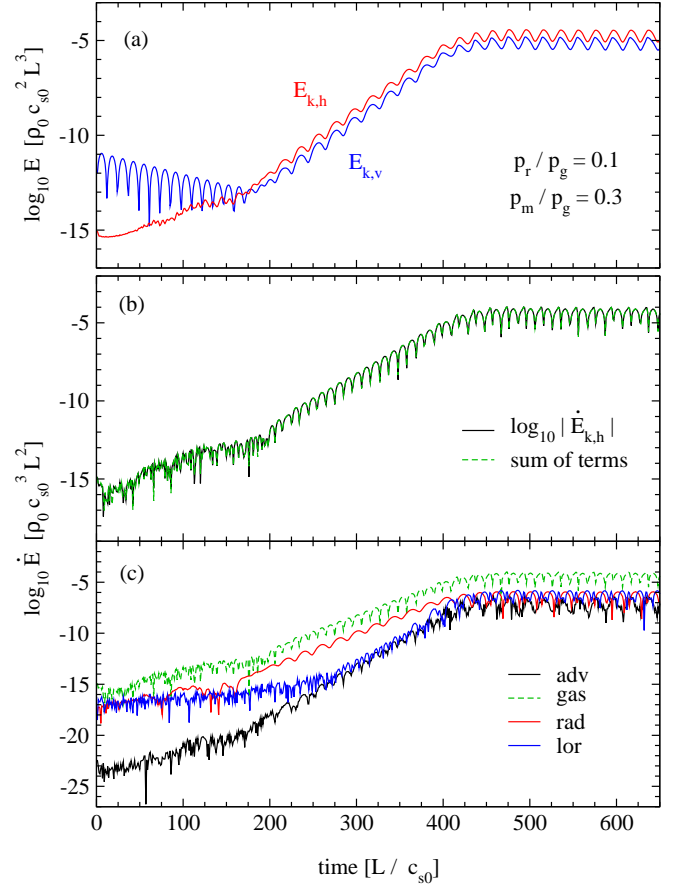}
\caption{\emph{Panel (a)}: Evolution of the volume-integrated vertical and horizontal kinetic energies 
(eqns.~[\ref{eq:Ekv_def}]-[\ref{eq:Ekh_def}]) as a function of time, for the model 
with $p_{\rm r}/p_{\rm g}=0.1$ and $p_{\rm m}/p_{\rm g}=0.3$ ($\theta_{\rm BF}=90^\circ$, resolution $128^2$). 
\emph{Panel (b):}
rate of change of the horizontal kinetic energy in the same model, obtained by differencing 
variables at consecutive time steps (black). Also shown is the sum of the terms that make up
the kinetic energy equation (green; eq.~\ref{eq:ekh_components}). \emph{Panel (c)} shows
each of these terms (eqns.~[\ref{eq:I_adv}]-[\ref{eq:I_lor}]) separately.}
\label{f:energy_evol_terms}
\end{figure}

The evolution of the volume-integrated kinetic energy for a gas-dominated 
model ($p_{\rm r}/p_{\rm g}=0.1$, $p_{\rm m}/p_{\rm g}=0.3$, $128^2$) is shown 
in Figure~\ref{f:energy_evol_terms}a, broken up into vertical 
and horizontal components,
\begin{eqnarray}
\label{eq:Ekv_def}
E_{k,v} & = & \int\left(\frac{1}{2}\rho\,v_z^2\right)\,\totd x\totd z\\
\label{eq:Ekh_def}
E_{k,h} & = & \int\left(\frac{1}{2}\rho\,v_x^2\right)\,\totd x\totd z\,
\end{eqnarray}
respectively. Initial density perturbations drive vertically-propagating waves that eventually decay. The horizontal
kinetic energy shows no such initial transients, growing out of numerical noise
from the beginning of the simulation. Both energies saturate at nearly same
time, closely following each other after the RMI overtakes the decaying waves.

The equation governing the horizontal kinetic energy is
\begin{eqnarray}
\label{eq:ekh_equation}
\frac{\partial}{\partial t}\left(\frac{1}{2} \rho v_x^2\right) & = & -\nabla\cdot \left[\rho\mathbf{v}\, \frac{1}{2}v_x^2 \right]
- v_x\frac{\partial}{\partial x}\left(p_{\rm g} + p_{\rm r} + p_{\rm m}\right)\nonumber\\
 && + \frac{v_x}{4\pi}\left(\mathbf{B}\cdot \nabla \right)B_x.
\end{eqnarray}
By itegrating over the computational volume,
we may write
\begin{equation}
\label{eq:ekh_components}
\dot{E}_{\rm k,h} = I_{\rm adv} + I_{\rm gas} + I_{\rm gas} + I_{\rm lor},
\end{equation}
where
\begin{eqnarray}
\label{eq:I_adv}
I_{\rm adv} & = & -\int\nabla\cdot \left(\rho\mathbf{v}\,\frac{1}{2}v_x^2 \right)\,\totd x\totd z\\
\label{eq:I_gas}
I_{\rm gas} & = & -\int v_x\frac{\partial p_{\rm g}}{\partial x}\,\totd x\totd z\\
\label{eq:I_rad}
I_{\rm rad} & = & -\int v_x\frac{\partial p_{\rm r}}{\partial x}\,\totd x\totd z\\
\label{eq:I_lor}
I_{\rm lor} & = & +\int \frac{v_x}{4\pi}\,B_z\left(\frac{\partial B_x}{\partial z}
		  -\frac{\partial B_z}{\partial x}\right)\,\totd x\totd z
\end{eqnarray}
are the contributions from advection, gas pressure gradients, radiation
pressure gradients, and the Lorentz force, respectively. We combine the contribution from
magnetic pressure and tension, since they cancel each other out to a large
degree due to their phase offset during linear growth.

Figure~\ref{f:energy_evol_terms}b shows the evolution of the rate of change of
the horizontal kinetic energy of the same model shown in panel (a). 
This rate is in excellent
agreement with the sum of the terms comprising
equation~(\ref{eq:ekh_components}), an additional verification of the 
accuracy of the solution method.

The terms that make up the rate of change of the horizontal kinetic energy are
shown separately in Figure~\ref{f:energy_evol_terms}c. The gas and radiation
pressure gradient terms grow at the same rate for $\sim 15$ e-folding times.
The advective and Lorentz force components grow more slowly initially, but then
increase their growth rate as the amplitude increases.

Saturation in this model occurs when the Lorentz force reaches the same
amplitude as the radiation pressure component. These two contributions are
$90^\circ$ out of phase, as expected from the form of the RMI forcing in the
linear regime (Appendix~\ref{s:linear_derivation}); the saturation
amplitude is small in this model ($\delta \rho/\rho\sim 3\times 10^{-2}$;
Table~\ref{t:results}). This result is consistent with that of \citet{turner2005},
who found that the RMI stopped growing when the field `buckled'.
The much lower radiative forcing in this model means that the amplitude
of magnetic field fluctuations required are much smaller.
This balance of stresses at saturation does not apply to other models with
lower $p_{\rm r}/p_{\rm g}$, however. A more careful study of this
interplay between magnetic and radiation stress tensors will be addressed
in future work.

\begin{figure}
\includegraphics*[width=\columnwidth]{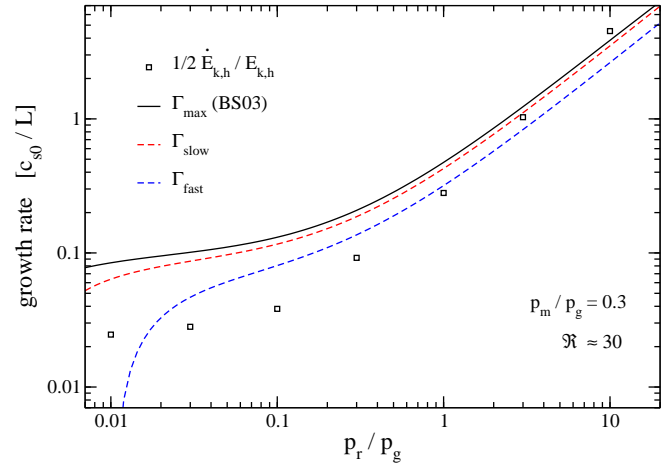}
\caption{Linear growth rates inferred from the evolution of the
horizontal kinetic energy as a function of the ratio of
radiation to gas pressure, for the sequence of models with fixed magnetic field (squares). 
The solid line shows the growth rate of the most unstable mode from the
dispersion relation of \citetalias{BS03}. 
The dashed lines are the approximate growth
rates of slow (red) and fast (blue) modes from equations~(\ref{eq:growth_slow}) 
and (\ref{eq:growth_fast}), respectively.}
\label{f:growth_parameter_space}
\end{figure}

The growth rates inferred from the evolution of $E_{\rm k,h}$
are shown in Figure~\ref{f:growth_parameter_space} for the
sequence of models with varying radiation pressure and fixed
magnetic field ($p_{\rm m}/p_{\rm g}=0.3$; $128^2$; $\theta_{\rm BF}=90^\circ$).
When radiation pressure is low, the measured values agree within a 
factor of a few with the growth rate of the most unstable mode 
from the dispersion relation of \citetalias{BS03}.
Agreement improves as radiation makes an 
increasingly larger contribution to the total pressure, and the
degree of forcing becomes larger. Comparing the 
growth rates measured in the model with $p_{\rm r}/p_{\rm g}=0.1$
for resolutions of $64^2$ and $128^2$ shows a decreasing 
agreement (at the few percent level) with coarser resolution,
pointing to numerical
dissipation as the likely cause of the discrepancy with the 
analytic value at low radiation pressure. The fact that numerical
growth rates exceed the analytic prediction at large radiation
pressure can perhaps be attributed to long-wavelength effects, as the \citetalias{BS03}
value is a correction of the order of $(kH_g)^{-1}$ to the magnetosonic frequency.
We nevertheless take these results as a confirmation of the predictions of \citetalias{BS03}
on the growth of the RMI when gas pressure dominates. 

\begin{figure*}
\includegraphics*[width=\textwidth]{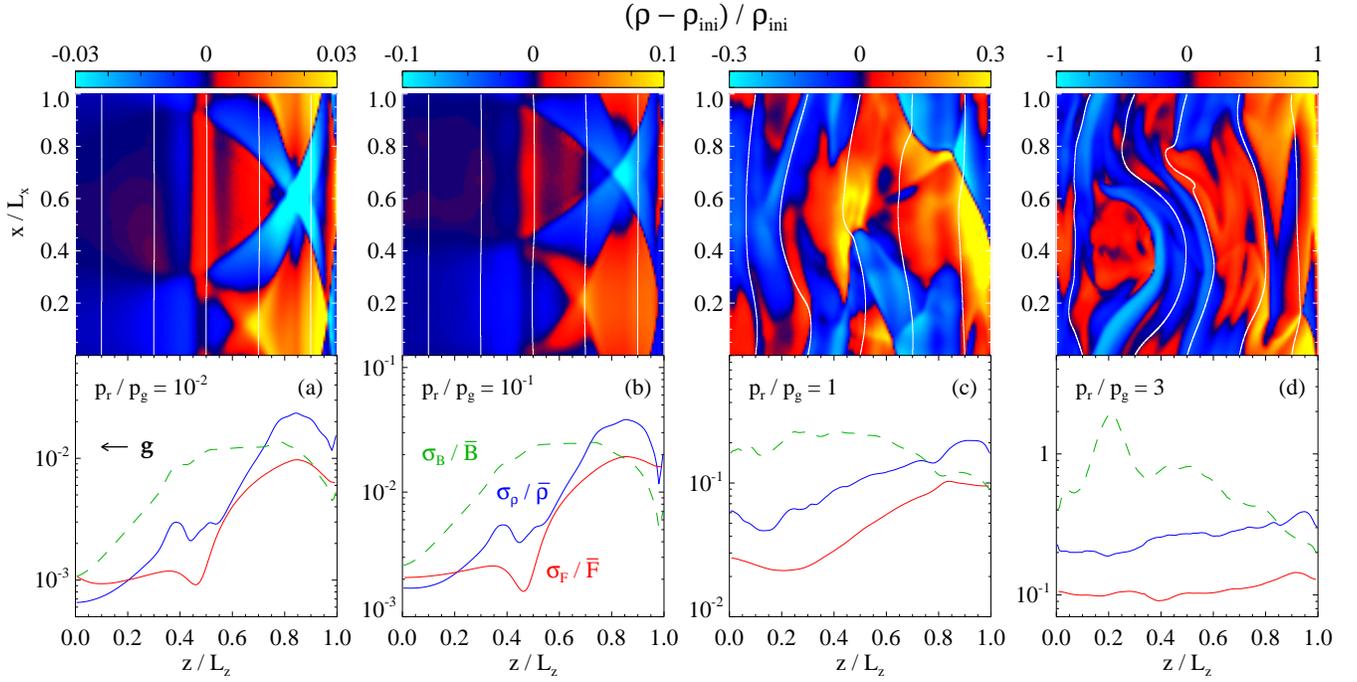}
\caption{\emph{Top:} Instantaneous deviation of the density relative to
its initial value  during saturation, for
models with increasing radiation pressure (as labeled) and constant magnetic field 
($p_{\rm m}/p_{\rm g} = 0.3$, resolution $128^2$, $\theta_{\rm BF}=90^\circ$). Curves show a few magnetic 
field lines. Note that gravity points in the negative $z$ direction.
\emph{Bottom:} Profiles of the time- and horizontal root-mean-square
fluctuation (eq.~[\ref{eq:rms_fluctuation}]) divided by mean 
value (eq.~[\ref{eq:mean_value}]) of density (blue), radiative flux (red), 
and magnetic field (green), for the same models shown in the top panels.}
\label{f:snapshots_prpg}
\end{figure*}

\begin{figure*}
\includegraphics*[width=\textwidth]{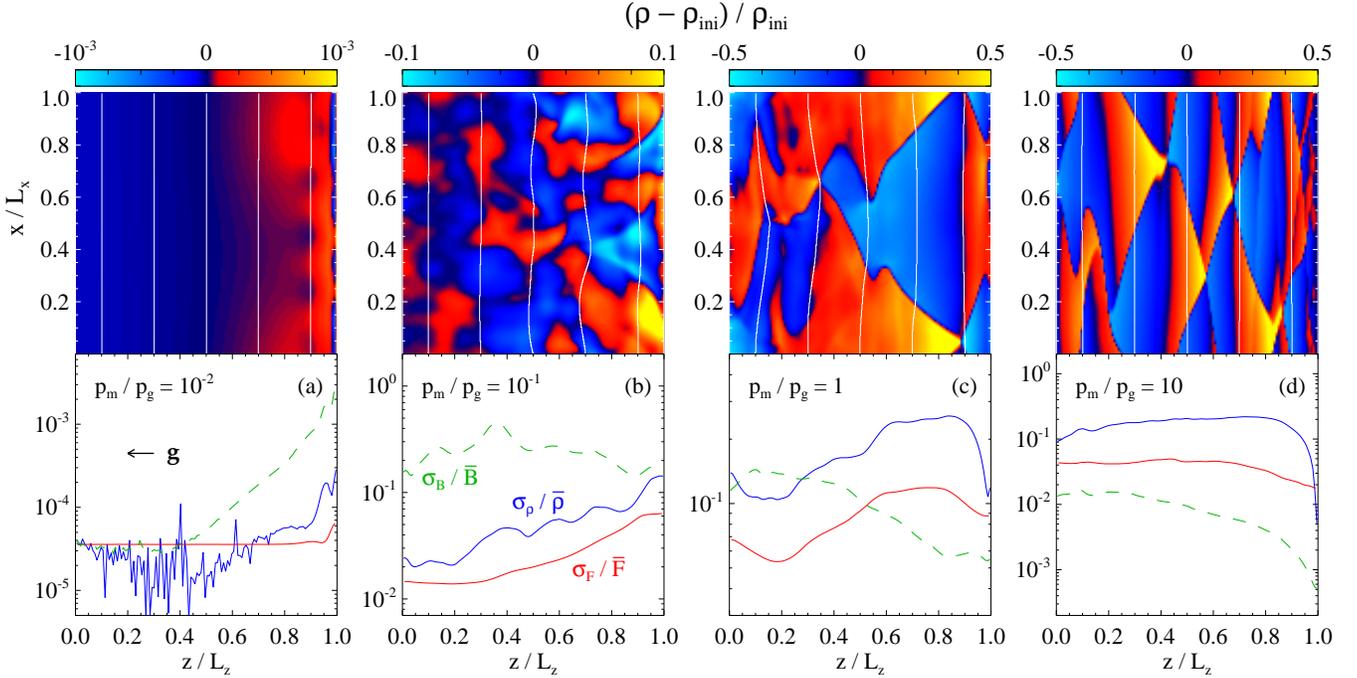}
\caption{Same as Figure~\ref{f:snapshots_prpg}, but for a range of models spanning different
ratios of magnetic pressure to gas pressure, as labeled. 
The radiation to gas pressure is fixed to $p_{\rm r}/p_{\rm g}=1$, and the resolution is $128^2$.
When the magnetic field dominates, the usual \emph{photon bubble} behavior is recovered (Panel d).}
\label{f:snapshots_pmpg}
\end{figure*}

Snapshots of the fractional change in density relative to the initial value during
saturation are shown in Figure~\ref{f:snapshots_prpg} for models with fixed
magnetic field and varying radiation to gas pressure ratios. When the radiation
pressure is low, amplitudes are small and the flow is dominated by domain-filling
structures. The magnetic field becomes significantly deformed when $p_{\rm r}\gtrsim p_{\rm m}$.
A larger contribution from radiation pressure increases the saturation amplitude and 
amount of small-scale structure.

The model with $p_{\rm r}\gtrsim p_{\rm m} = 10$ and $p_{\rm m}/p_{\rm g}=0.3$ 
deserves special attention.
The amplitude of the fluctuations becomes large enough that
the diffusion solver fails to converge. Future studies should address this region of
parameter space using a suitable closure for the radiation moment equations and a 
larger computational domain.

Another set of snapshots is shown in Figure~\ref{f:snapshots_pmpg} for models
with fixed $p_{\rm r}/p_{\rm g}$ and increasing magnetic field. Despite the fact
that the contribution from radiation is significant, very small amplitudes are
obtained when magnetic fields are low. In fact, no growth is seen at all
for the lowest field configuration when the resolution is $32^2$, in contrast to
most other models. Decreasing the radiation pressure to $p_{\rm r}/p_{\rm g}=0.3$
and keeping $p_{\rm m}/p_{\rm g}=0.01$ leads to no growth even at our baseline $128^2$
resolution.
Increasing the magnetic field strength increases the amplitude
and spatial size of the dominant structures up to the point where $p_{\rm m} \sim p_{\rm r}$.
Further increases in magnetic field cause the saturation amplitude to decrease slightly,
giving rise to the structures seen by \citet{turner2005}.

\subsection{Time-Averaged Properties}
\label{s:time_averaged}

We now discuss the quantitative behavior of the simulations
in the saturated state. Saturation is defined 
as a period during which the kinetic
energy is constant when averaged over timescales longer
than the oscillation period or growth time. In some cases
(e.g., $p_{\rm r}/p_{\rm g}= 1$), the system
achieves an initial phase of saturation lasting a few 
hundred sound crossing times, but then mass begins to
leave the box at an increasingly rapid rate, changing its
global properties. We always choose the time interval
for averaging such that the mass in the box remains nearly constant.

\begin{figure}
\includegraphics*[width=\columnwidth]{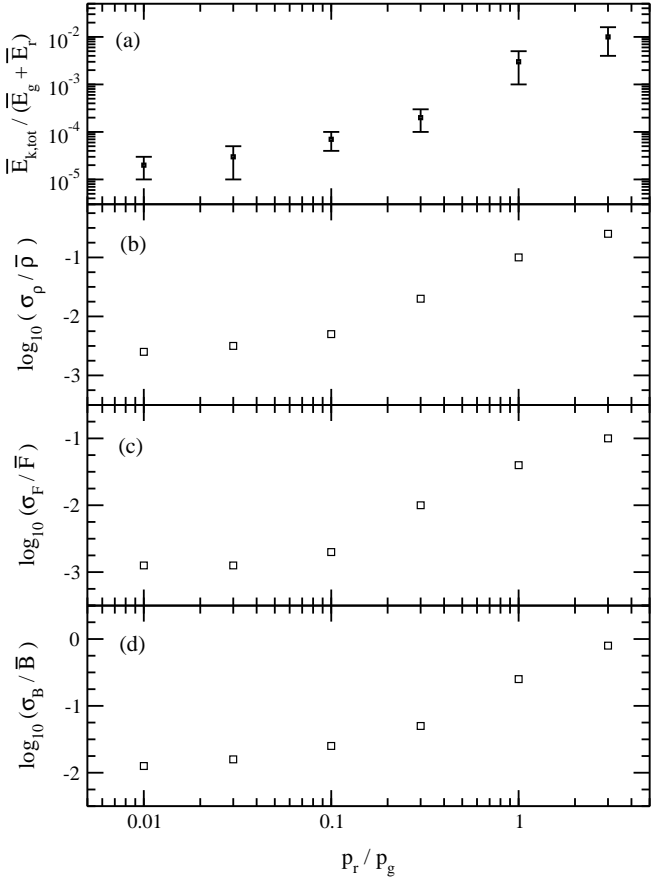}
\caption{Time-averaged properties as a function of the ratio of radiation to gas pressure, for
models with $p_{\rm m}/p_{\rm g} = 0.3$ ($\theta_{\rm BF}=90^\circ$, resolution $128^2$). 
\emph{Panel (a)} shows the ratio of total kinetic energy to the sum of the gas and 
radiation energies. Error bars show the size of the corresponding root mean square fluctuation. 
\emph{Panels (b), (c), and (d)} show the ratio of the root mean square fluctuation to mean 
value at domain center of the density, vertical radiative flux, and magnetic field strength, respectively.}
\label{f:timeave_prpg}
\end{figure}

The time-average of the volume-integrated vertical kinetic-, horizontal kinetic-, 
gas-, and radiation energies is denoted by 
$\bar{E}_{\rm k,v}$, $\bar{E}_{\rm k,h}$, $\bar{E}_g$, and $\bar{E}_r$, respectively.
For field variables, 
we compute time- and horizontal averages
\begin{equation}
\label{eq:mean_value}
\overline A = \frac{1}{(T\,L_x)}\int \totd t\int \totd x\, A,
\end{equation}
where $A(x,z,t)$ stands for a scalar such as density.
The root-mean-square fluctuation is computed as
\begin{equation}
\label{eq:rms_fluctuation}
\sigma_{\rm A} = \left[ \overline{A^2} - \left(\overline{A} \right)^2\right]^{1/2}.
\end{equation}

For all the models that reach saturation, we report in Table~\ref{t:results}
the ratio of total time-average
kinetic energy to time-averaged internal plus radiation energies 
$\bar{E}_{\rm k,tot}/(\bar{E}_g+\bar{E}_r)$, where $\bar{E}_{\rm k,tot}=\bar{E}_{k,v}+\bar{E}_{k,h}$,
and the ratio of vertical to horizontal kinetic energies $\bar{E}_{k,v}/\bar{E}_{k,h}$. 
In addition, ratios of root-mean-square fluctuation to mean value of density
$\sigma_\rho / \bar{\rho}$, vertical component of the radiative flux $\sigma_{\rm F}/\bar{F}$,
and magnetic field strength $\sigma_{\rm B} /\bar{B}$ are provided at three
points in the computational box. The lower panels in 
Figures~\ref{f:snapshots_prpg} and \ref{f:snapshots_pmpg} show
that these averaged variables change considerably with altitude, sometimes by up to an 
order of magnitude. To capture this variation, values at $20\%$, $50\%$, and $80\%$ 
of the box height are shown in Table~\ref{t:results}.

Figure~\ref{f:timeave_prpg} shows the kinetic energy as a function
$p_{\rm r}/p_{\rm g}$, for the sequence of runs with constant
$p_{\rm m}/p_{\rm g}=0.3$ and resolution $128^2$.
The kinetic energy is a monotonic function of the ratio of radiation 
to gas pressure, increasing the fastest at $p_{\rm r}/p_{\rm g}\sim 0.5$. 
A similar trend is observed for the root-mean-square
fluctuation of the density, radiative flux, and magnetic field strength, whose
values at domain center are also shown in the same figure.
Such increase in steady-state amplitudes is clearly 
the consequence of the fact that the radiation field is the 
ultimate source of energy (\S\ref{s:aspects}).
The amplitude of the fluctuations in radiative flux and density are comparable to each other,
while the magnetic field perturbations are larger by an order of magnitude at box center.
This discrepancy is in part a consequence of the different spatial distributions of 
fluctuations in the box (e.g., lower panels of Figure~\ref{f:snapshots_prpg}).

\begin{figure}
\includegraphics*[width=\columnwidth]{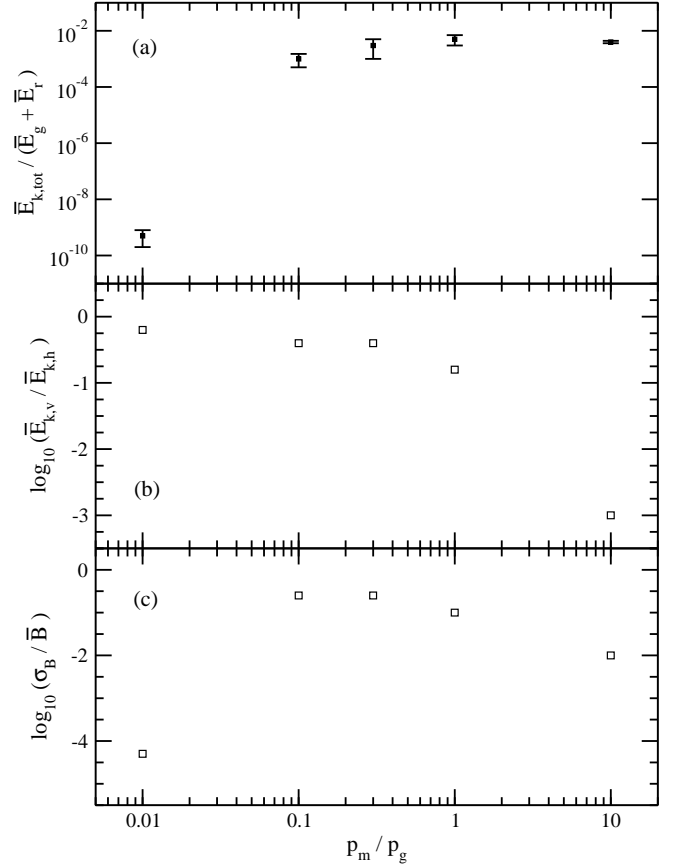}
\caption{Time-averaged properties as a function of the ratio of magnetic to gas pressure, 
for models with $p_{\rm r}/p_{\rm g}=1$ ($\theta_{\rm BF}=90^\circ$, resolution $128^2$).
\emph{Panel (a)} shows the ratio of the total kinetic energy to the sum of the gas 
and radiation energies. The ratio of the vertical to horizontal kinetic energies
is shown in \emph{panel (b)}, and the normalized root mean square fluctuation at domain center
of the magnetic field strength in \emph{panel (c)}.}
\label{f:timeave_pmpg}
\end{figure}

The dependence of the saturated quantities on the ratio of magnetic to gas
pressure is shown in Figure~\ref{f:timeave_pmpg} for the sequence of simulations
with $p_{\rm r}/p_{\rm g}=1$. As pointed out in \S\ref{s:evolution},
the kinetic energy increases with field strength when the inequality $p_{\rm m}\ll p_{\rm r}\sim p_{\rm g}$
obtains. Maximal amplitude is obtained when the contributions to the pressure are all
comparable to each other. 
Further increase in the magnetic field suppresses vertical motions,
leading to a lower total kinetic energy. This is clearly shown by the sharp decrease
in the ratio of vertical to horizontal kinetic energy when going from $p_{\rm m}/p_{\rm g}=1$ to $10$
in Figure~\ref{f:timeave_pmpg}b.
The amplitude of the magnetic field fluctuations follow the magnitude of the kinetic energy.

Table~\ref{t:results} also shows that varying the inclination of the magnetic field
relative to gravity changes the saturation amplitude. The optimal configuration
appears to be a field perpendicular to gravity. When the magnetic field comes close
to the vertical direction, the growth rates are eventually suppressed by the sine
of the angle between field and radiative flux (equation~\ref{eq:growth_rate_first_order}).
This change of saturation amplitude with angle was also seen by \citet{turner2005}.

\subsection{Dependence on Numerical Parameters}
\label{s:numerical_parameters}

\begin{figure}
\includegraphics*[width=\columnwidth]{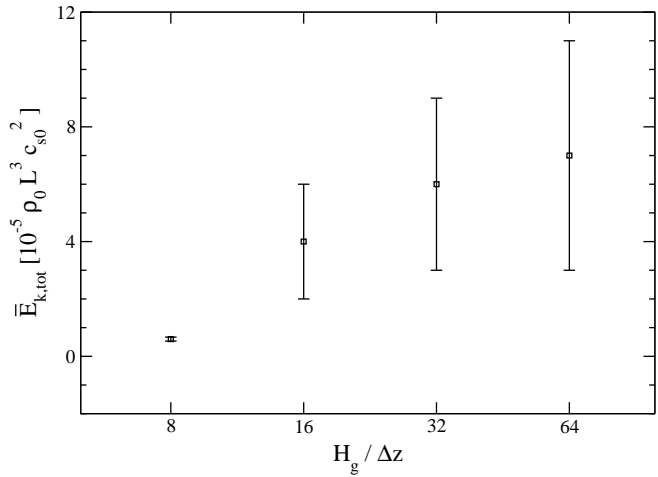}
\caption{Time-averaged kinetic energy as a function of spatial resolution.
Parameters of the model are $p_{\rm r}/p_{\rm g}=0.1$, $p_{\rm m}/p_{\rm g}=0.3$, and $\theta_{\rm BF}=90^\circ$. 
Error bars denote the root mean square fluctuation.}
\label{f:ekin_resol}
\end{figure}

We have selected a gas-dominated model ($p_{\rm r}/p_{\rm g}=0.1$, $p_{\rm r}/p_{\rm g}=0.3$, $\theta_{\rm BF}=90^\circ$)
for studying the dependence of the saturation properties on the choice of numerical parameters.
Figure~\ref{f:ekin_resol} shows that the time-averaged kinetic energy in the box as a function
of resolution is approaching convergence when using $64$ cells per gas pressure scale
height in each direction ($128^2$ cells in the box). The difference with using half the 
number of cells per direction decreases the kinetic energy by $\sim 15\%$. We thus consider
our simulations are spatially well resolved. Note that using less than $10$ cells per
gas pressure scale height can significantly underestimate the saturation amplitude.

\citet{turner2005} found that the choice of time step is also relevant for correctly
capturing the growth of the instability when using an implicit solver for the radiation sector.
We have performed an additional set of simulations (not shown in Table~\ref{t:results})
that vary the time-step employed on the model with $64^2$ resolution. No appreciable
difference in the saturation amplitude is found
when increasing the minimum time step from the fiducial value, which resolves diffusion
over a wavelength $H_{\rm g}/16$ (\S\ref{s:model_parameters}),
to the diffusion time for a wavelength $H_{\rm g}/4$ . Minor differences
of the order of $0.1$ dex in the amplitude of fluctuations of field variables are the only
difference. One possible cause of this is that the nonlinear phase is dominated by
large scale structures that are well resolved.

The size of the box in the horizontal direction can change the value of the saturation
amplitude by a factor of order unity. Table~\ref{t:results} shows that the simulation
with $64^2$ resolution has less than half the kinetic energy (relative to the sum of
the gas and radiation energies) of that where the horizontal
size is $4H_{\rm g}$ ($128\times 64$). 
In both cases, the dominant mode fills the domain laterally (as in Figure~\ref{f:snapshots_prpg}b),
and the mode with longer wavelength seems to extract more power from the radiation field.
Further increasing the box size to $8H_{\rm g}$ causes the relative kinetic energy to 
decrease. The dominant mode in this case has a lateral wavelength
of $8H_{\rm g}/3$. The relative kinetic energy is larger than that of the
model with a side of $2H_{\rm g}$, but less than that of the model with a side of $4H_{\rm g}$,
in agreement with a trend of monotonically increasing power with dominant mode wavelength.
Further work should investigate the origin of the dominant mode wavelength, when
periodic boundary conditions are used.

\section{Application: Massive Stars}
\label{s:implications}

Massive main-sequence and blue supergiant stars have radiative envelopes, with
an important contribution from radiation pressure to the
overall support. Many OB stars have recently been found
to have magnetic fields up to kG strength or even higher
(see, e.g., \citealt{petit2012} for a recent compilation).
Conditions near the surface of these stars are clearly 
susceptible to the RMI\footnote{%
Following \citetalias{BS03}, our analysis assumes a constant magnetic field on length
scales comparable to the gas pressure scale height.
This is a good starting point to address the prevalence of this instability
in stars with radiative envelopes, though it is not necessarily a realistic approximation.
}.

The winds of massive stars are known to be time-variable and show
clumpiness over a wide range of spatial scales. 
For instance, mass loss rates inferred from
UV resonance lines -- which scale linearly with density -- yield
estimates that are at least an order of magnitude lower than
those from collisional emission processes (e.g., H$\alpha$), 
which depend on the density squared \citep{fullerton2006}. 

Variability phenomena are traditionally divided into 
large- and small-scale. Below we elaborate on the potential
impact of the RMI on each of these two categories.

\subsection{Large-Scale Variability}

It is currently thought that 
cyclic variability on scales comparable to the stellar radius
(e.g., time-dependent absorption superimposed on the blue
edge of P Cygni profiles) originates in so-called 
Corotating Interacting Regions (CIRs), in analogy with the Solar Wind 
\citep{mullan1984}. 
In this model, azimuthal fluctuations at the 
base of the wind cause faster streams to overtake slower ones, generating 
shocks and variations in the optical depth that follow a spiral 
pattern due to angular-momentum conservation (e.g., \citealt{cranmer1996}, 
\citealt{dessart2004}).

The source of these fluctuations in massive stars is
not established yet. Proposed mechanisms include flow in a 
spatially inhomogeneous magnetic field \citep{underhill1984} and multi-periodic
non-radial pulsations \citep{kaufer2006}, the latter possibly excited by 
near surface convection zones from opacity peaks \citep{cantiello2009}. 
In at least one source (HD 64760), 
the spectral line variability can be accounted for if the spots do not
corotate with the stellar surface \citep{lobel2008}, lending support to the multi-periodic
non-radial pulsation hypothesis.  From our results, the RMI is a clear candidate 
as a source of such pulsations.

\subsection{Small-Scale Variability}

Line-driven winds are subject to the Line Deshadowing Instability (LDI) above
the photosphere \citep{lucy1970,macgregor1979,owocki1984}. When
the velocity of a parcel of gas in the wind is perturbed outward, the corresponding
Doppler shift moves the relevant rest-frame transitions into an energy range
where the continuum flux from the star is higher, increasing the outward acceleration.
The instability tends to favor small spatial scales. The nonlinear development of 
the LDI results in extended regions of sharp density and velocity fluctuations that
lead to the compression of wind material into thin radial shells \citep{owocki1988}.
These shells are subsequently broken up by secondary instabilities,
resulting in a clumped wind \citep{dessart2003}. It is currently thought that the LDI
can account for most of the phenomenology requiring small scale inhomogeneities,
such as transient emission line substructures, black troughs of saturated UV resonance
lines, discrepant mass-loss estimates, and X-ray emission (e.g., \citealt{fullerton2011}).

A remaining question is the presence of clumping at the base of the wind. 
Simulations of the `self-excited' LDI find that damping due to scattered radiation
\citep{lucy1984} suppresses the onset of fluctuations below $\sim 1.5$ stellar radii \citep{runacres2002}. 
A number of observational studies have found evidence for clumping much closer to the base of the wind.
In particular, \citet{najarro2011} used different line diagnostics to reconstruct the
radial profile of the clumping factor (average of $\rho^2$ over the squared average of $\rho$)
for the entire wind inside $\sim 100$ stellar radii in $\zeta$ Pup, finding that the \emph{peak} of this factor 
occurs inside $2$ stellar radii. Recently, \citet{sundqvist2012} have reported LDI simulations
that agree with the \citet{najarro2011} clumping profile when including limb-darkening effects (which decrease
line-drag from scattered radiation) and generic sound wave perturbations  
at the base of the wind.

The RMI can readily drive sub-photospheric density perturbations on small spatial scales.
\citetalias{BS03} found that in the thermally locked
limit (eq.~[\ref{eq:thermal_locking_condition}]) the instability operates
up to a wave number
\begin{equation}
k_{\rm cut} = \left\{
\begin{array}{cc}
\left(\zeta^{-5}\,\omega_{\rm th}\, g/c_i^3\right)^{1/2} & \qquad\textrm{(slow)}\\
\noalign{\medskip}
\left(\zeta^5\,\omega_{\rm th}\, g /v_{\rm A}^3\right)^{1/2} & \qquad\textrm{(fast)}
\end{array}
\right.,
\end{equation}
where $\zeta$ is given by equation~(\ref{eq:zeta_mag}). For a near equipartition
magnetic field, this translates into a minimum unstable wavelength of
\begin{equation}
\lambda_{\rm cut}\sim 6\,(\rho_{-8} \,g_3)^{-1/2}\,c_{\rm i,6}^{3/2}\,
      		      \left(\frac{p_{\rm g}}{p_{\rm r}} \right)^{1/2} 
		      \left(\frac{\kappa_{\rm es}}{\kappa_a} \right)^{1/2}\,\textrm{ km},
\end{equation}
where $\rho_{-8}$ is the density in units of $10^{-8}$~g~cm$^{-3}$, $g_3$ the acceleration
of gravity in units of $10^3$~cm~s$^{-2}$, $c_{i,6}$ the isothermal sound speed in units
of $10$~km~s$^{-1}$, and $\kappa_{\rm es}$ is the electron scattering opacity. Comparing
to the gas pressure scale height at the photosphere,
\begin{eqnarray}
H_{\rm g,ph} & \sim & (\kappa_{\rm F}\rho_{\rm ph})^{-1}\\
	     & \sim & 3000\rho_{-8}^{-1}\left(\frac{\kappa_{\rm es}}{\kappa_{\rm F}} \right)\textrm{ km},
\end{eqnarray}
shows that the RMI in its WKB version operates over a dynamic range of at 
least $100$ in wavelength. 

\section{Summary}
\label{s:summary}

In this paper we have studied the nonlinear development
of the Radiation-Driven Magneto-Acoustic Instability (RMI)
in the regime where the gas pressure is comparable or 
dominates over that due to the radiation and magnetic field, 
and in the limit
of rapid heat exchange between matter and radiation (eq.~[\ref{eq:thermal_locking_condition}]). 
Our primary results are:
\newline

\noindent 1. -- We have confirmed the predictions of Blaes and Socrates (2003).
		In particular, the RMI instability operates when gas
		pressure dominates over radiation pressure and when the magnetic
		field is weak. 
		The growth rate of the kinetic energy is consistent with that
		expected for the most unstable mode
		(Fig.~\ref{f:growth_parameter_space}).
		\newline

\noindent 2. -- The saturation amplitude is an increasing function of the ratio of
		radiation to gas pressure (Fig.~\ref{f:timeave_prpg}) over all of the
		dynamic range studied here, $p_{\rm r}/p_{\rm g} \in [10^{-2},10]$. 
	 	At the lowest end of this interval, the saturation amplitude is
		small but non-negligible ($\delta\rho/\rho \sim 10^{-3}$).
		\newline

\noindent 3. -- The saturated state varies non-monotonically with magnetic field
		strength (Fig.~\ref{f:timeave_pmpg}). For moderate forcing by radiation ($p_{\rm r}\sim p_{\rm g}$),
		the saturation amplitude increases with field strength as long as
	        magnetic pressure is sub-dominant. Once the magnetic field dominates,
		vertical motions are suppressed and increasingly smaller amplitudes are obtained.
		\newline

\noindent 4. -- The RMI may be a source of sub-photospheric perturbations
		in the envelopes of massive main-sequence and blue supergiant stars.
		The instability can serve as a source of large spatial scale 
		perturbations for Corotating Interacting Regions.
		The RMI also operates over a wide dynamic range in wavelength,
		providing sufficient seed perturbations for the Line Deshadowing 
		Instability even if a convection zone is absent. 
		\newline
	
The computational requirements to capture the RMI in realistic
contexts are very demanding. The instability is strongest in 
regions near the photosphere, where diffusion is
fast relative to the mode period and where the pressure scale
height is much smaller than the stellar radius. 
Capturing the instability requires resolving diffusion over the pressure scale 
height in space and in time, introducing an additional computational cost. 
For example, the ratio of the sound crossing time
over the diffusion time at a given wavelength is proportional
to the diffusion speed over the sound speed ($\propto \Re$, eq.~[\ref{eq:rapid_diffusion_parameter}]). 
For massive stars this ratio can be as high as $\sim 10^{5}$, requiring
a comparable number of diffusion steps per Courant time to capture the RMI.

Future work will address the saturation of the RMI, 
and further application to astrophysical systems.

\acknowledgements
We thank the anonymous referee for helpful comments that improved the manuscript.
RF is supported in part by NASA through Einstein Postdoctoral Fellowship
grant number PF-00062, awarded by the Chandra X-ray Center, which is operated
by the Smithsonian Astrophysical Observatory for NASA under contract NAS8-03060,
and by NSF grant number AST-0807444.
AS acknowledges support from a John N. Bahcall Fellowship, awarded by the Institute for
Advanced Study, Princeton.
This research was also supported in part by the National Science Foundation through
TeraGrid/XSEDE resources \citep{catlett07}. Computations were performed at
LONI Queen Bee and the IAS Aurora cluster.

\appendix

\section{Derivation of Linear Growth Rates and Stability Criteria}
\label{s:linear_derivation}

\subsection{Basic Magneto-Acoustic Waves}

In the absence of stratification and a changing radiation field, compressive fluid perturbations
are forced by a combination
of their own acoustic response and the Lorentz force. 
In this limit, conservation of momentum to linear order can be written as
\begin{equation}
\label{eq:momentum_cons_linear}
-i\,\omega_0\rho\,\delta{\bf v}=-i\,{\bf k}\left[ \delta p_{\rm g}+\frac{{\bf B}\cdot\delta{\bf B}}{4\pi}   \right]
+i\frac{{\bf k}\cdot{\bf B}}{4\pi}\delta{\bf B}
\end{equation}
where $\delta$ denotes Eulerian perturbation, and $\delta p_{\rm g} =  c^2_i\delta\rho$ when
$\delta T \to 0$ (eq.~[\ref{eq:thermal_locking_condition}]).  We assume
that all perturbed quantities take plane-wave form and are 
$\propto \exp\left({\bf k}\cdot {\bf x}-\omega t\right)$.
Using the linearized mass conservation equation,
the linearized induction equation
   \begin{equation}
   \label{eq:induction_linear}
   \delta{\bf B}=\frac{{\bf B}}{\omega_0}\left({\bf k}\cdot\delta{\bf v}\right)-\frac{\left({\bf k}\cdot{\bf B}\right)}{\omega_0}
   \,\delta{\bf v},
   \end{equation}
and the component of the linearized momentum equation along ${\mathbf B}$,
one can rewrite equation (\ref{eq:momentum_cons_linear}) as
\begin{equation}
\label{eq:momentum_cons_linear2}
\frac{\tilde{\omega}^2}{\omega_0}\delta{\bf v}=
                 \frac{\delta \rho}{\rho}\left[\left( \frac{\tilde\omega^2}{\omega_0^2}\,c_i^2
            +v_{\rm A}^2\right)\mathbf{k} -\left({\bf k}\cdot{\bf v}_{\rm A}\right)\,\mathbf{v_{\rm A}}\right],
\end{equation}
where $\tilde\omega^2 = \omega_0^2 - (\mathbf{k}\cdot\mathbf{v}_{\rm A})^2$.
Projecting onto $\mathbf{k}$ and using mass conservation yields the
magnetosonic dispersion relation,
\begin{eqnarray}
\label{eq:disprel_magnetosonic}
v_{\rm ph}^2 & \equiv & \frac{\omega_0^2}{k^2} = \frac{\tilde{\omega}^2}{\omega_0^2}\,c_i^2 + v_{\rm A}^2\\
 & = & \frac{1}{2}\left[(c_{\rm i}^2+v_{\rm A}^2)\pm \sqrt{(c_{\rm i}^2+v_{\rm A}^2)^2
-4(\hat k\cdot \mathbf{v}_{\rm A})^2 c_{\rm i}^2} \right].
\end{eqnarray}

\subsection{Radiative Correction: Driving and Diffusive Damping}

We now include stratification in the background state and allow the radiation field (or equivalently
the temperature) to vary. The exchange of heat between radiation and energy is
treated as a correction to the mode frequency.
We thus have
\begin{equation}
\label{eq:omega_complex_definition}
\omega =  \omega_0 + i\Gamma,
\end{equation}
where $\Gamma$ is a driving or damping rate such that $\Gamma \ll \omega_0$.
The linearized momentum equation is now
\begin{equation}
-\omega^2\,\rho{\bm\xi} = -i\,{\bf k}\left[ \delta p_{\rm g} + \delta p_{\rm r}
        +\frac{{\bf B}\cdot\delta{\bf B}}{4\pi} \right] +i\frac{{\bf k}\cdot{\bf B}}{4\pi}\delta{\bf B} + \mathbf{g}\delta\rho,
\end{equation}
where $\bm \xi = i\delta\mathbf{v}/\omega$ is the Lagrangian displacement, and
$p_{\rm r} = E/3 = aT^4/3$ is the radiation pressure in the optically thick limit.

Defining $P\equiv p_{\rm g}+ p_{\rm r}$, using the condition of hydrostatic equilibrium
$\rho\mathbf{g} = \nabla P$, and expressing Eulerian perturbations in
terms of the Lagrangian displacement
\begin{eqnarray}
\delta\rho       & = & -i(\mathbf{k}\cdot{\bm\xi}) -{\bm\xi}\cdot\nabla\rho\\
\delta P         & = & c_i^2\delta\rho + \left(\frac{\partial P}{\partial T} \right)_\rho\, \delta T\\
\delta\mathbf{B} & = & i(\mathbf{k}\cdot\mathbf{B}){\bm\xi}-i(\mathbf{k}\cdot{\bm \xi})\mathbf{B},
\end{eqnarray}
one can re-express momentum balance as
\begin{equation}
\label{eq:eom_force_lagrangian}
\left(-\omega^2 + \omega_0^2 \right){\bm\xi} = \delta {\bm {\mathcal{F}}} + i\,c_i^2{\bm \xi}\times (\mathbf{k}\times \nabla\ln\rho)
\end{equation}
where
\begin{equation}
\label{eq:forcing_definition}
\delta {\bm {\mathcal{F}}}=
-\frac{i}{\rho}\left(\frac{\partial P}{\partial T}\right)_{\rho}\left[{\bf k}\delta T
+\left({\bf k}\cdot{\bm\xi}\right)\nabla T \right]
\end{equation}
contains the driving and damping terms. The second term on the left hand side of equation~(\ref{eq:eom_force_lagrangian})
includes the terms arising from acoustic and magnetic forces (equation~\ref{eq:momentum_cons_linear2}).
The second term on the right hand side is perpendicular to the Lagrangian displacement and
hence cannot do work on the perturbation (e.g., \citealt{socrates2005}). Terms of higher order in $(k H_{\rm g})^{-1}$
have been discarded.  

To first order in $\Gamma/\omega_0$, the growth rate is
\begin{equation}
\label{eq:work_integral}
\Gamma \simeq \frac{1}{2}\frac{{\bm {\mathcal{F}}}\cdot \delta\mathbf{v}}{\delta\mathbf{v}\cdot\delta\mathbf{v}}.
\end{equation}
Before evaluating equation~(\ref{eq:work_integral}),
it is instructive to explicitly separate the driving and damping terms in equation~(\ref{eq:forcing_definition}).
To do so we solve for the temperature perturbation in the linearized equation for the total energy density.
Defining $U\equiv e + E$, we have
\begin{equation}
-i\omega\delta U + \delta\mathbf{v}\cdot\nabla U + i(U+P)\mathbf{k}\cdot\delta\mathbf{v} = -i\mathbf{k}\cdot\delta\mathbf{F}
\end{equation}
or
\begin{eqnarray}
\label{eq:perturbed_energy_2}
-i\omega\left[\left( \frac{\partial U}{\partial T}\right)_\rho\delta T +
                     nT\left(\frac{\partial s}{\partial \rho} \right)_T\delta\rho\right]
 + nT\delta\mathbf{v}\cdot \nabla s = \nonumber\\
i\left(\mathbf{k}\cdot \mathbf{F}\right)\,\frac{\delta\rho}{\rho}
-\omega_{\rm diff}\left(\frac{\partial E}{\partial T} \right)_\rho\delta T,
\end{eqnarray}
where
\begin{equation}
\label{eq:dsdrho}
\label{eq:silk_damping}
nT\left(\frac{\partial s}{\partial\rho} \right)_T = \left(\frac{\partial U}{\partial \rho}\right)_T - \frac{U+P}{\rho}
= -\frac{1}{\rho}\left(p_{\rm g} + \frac{4}{3}E \right).
\end{equation}
To first order in $\omega/\omega_{\rm diff}$
and in $(kH_{\rm g})^{-1}$, the temperature perturbation is
\begin{equation}
\label{eq:dT_driving_damp}
\frac{\delta T}{T} \simeq \frac{i}{4}\left[
\frac{\omega}{\omega_{\rm diff}}\frac{nT}{E}\left(\frac{\partial s}{\partial\ln\rho} \right)_T
+\frac{\mathbf{k}\cdot \mathbf{F}}{\omega_{\rm diff}E}
\right]\,
\frac{\delta \rho}{\rho}.
\end{equation}
The first term represents damping by radiative diffusion (eq.~[\ref{eq:dsdrho}]):
entropy is lost upon compression\footnote{In
cosmology, this effect is known as ``Silk damping", after \citet{silk1968}.}.
Note that we have dropped the first term  on the left hand side of
equation~(\ref{eq:perturbed_energy_2}); it becomes important if
$(p_{\rm r}/p_{\rm g})\lesssim (\omega/\omega_{\rm diff})$, contributing with a global factor of
order unity to the growth rate (e.g., compare $\Gamma_{\rm slow}$ and the maximum growth rate
from the dispersion relation in Figure~\ref{f:growth_parameter_space}; the choice of parameters
is such that $\omega_0/\omega_{\rm diff} = 0.01$).

We now evaluate equation~(\ref{eq:work_integral}) by 
inserting equation~(\ref{eq:dT_driving_damp}) into equation~(\ref{eq:forcing_definition}), projecting
onto $\delta\mathbf{v}$, using equation~(\ref{eq:momentum_cons_linear2}) to eliminate a term
proportional to $(\mathbf{k}\times \delta\mathbf{v})$, and keeping the same order of approximation,
we find
\begin{eqnarray}
\label{eq:Fdotdv}
{\bm {\mathcal{F}}}\cdot \delta\mathbf{v} = &&\left(\frac{\delta\rho}{\rho} \right)^2\,
\frac{(\partial P / \partial T)_\rho}{(\partial E / \partial T)_\rho}\,
\left(\frac{\omega}{\omega_{\rm diff}} \right)\omega\,\times\nonumber\\
&&\frac{1}{\rho}\left[nT\left(\frac{\partial s}{\partial \ln \rho} \right)_T +
\frac{(\mathbf{k}\cdot\mathbf{v}_{\rm A})\cdot(\mathbf{k}\times \mathbf{v}_{\rm A})\cdot(\mathbf{k}\times\mathbf{F})}
{\tilde\omega^2\,\omega} \right].\nonumber\\
\end{eqnarray}
The second term inside the square brackets represents driving by the background radiative flux (\citetalias{BS03}).
Note that only the component of the flux perpendicular to $\mathbf{k}$ contributes to the driving,
and that driving vanishes when the magnetic field is completely parallel or perpendicular to
the wave vector. By using equation~(\ref{eq:momentum_cons_linear2}), the denominator of equation~(\ref{eq:work_integral})
can be written as
\begin{equation}
\label{eq:delta_v_squared}
\delta\mathbf{v}\cdot\delta\mathbf{v} = \left(\frac{\delta\rho}{\rho}\right)^2\,\frac{\omega^2}{\tilde\omega^2}\,
\left(2v_{\rm ph}^2 - c_{\rm i}^2 - v_{\rm A}^2 \right).
\end{equation}
Equations~(\ref{eq:omega_diff}), (\ref{eq:work_integral}),
(\ref{eq:Fdotdv}), (\ref{eq:silk_damping}) and (\ref{eq:delta_v_squared})
then yield the growth rate to first order in
$(\Gamma/\omega_0)$, $(\omega_0/\omega_{\rm diff})$, and $(kH_{\rm g})^{-1}$,
\begin{eqnarray}
\label{eq:growth_rate_first_order}
\Gamma \simeq &&\frac{1}{2} \left[\frac{v_{\rm ph}^2 -(\hat k\cdot \mathbf{v}_{\rm A})^2}
{2v_{\rm ph}^2 - c_{\rm i}^2 - v_{\rm A}^2}\right]\left(\frac{3p_{\rm g}}{4E} + 1 \right)\times\nonumber\\
&&\frac{\kappa_{\rm F}}{c v_{\rm ph}}
\left[\frac{(\hat k\cdot \mathbf{v}_{\rm A})(\hat k\times \mathbf{v}_{\rm A})(\hat k\times \mathbf{F})}
           {[v_{\rm ph}^2-(\hat k\cdot \mathbf{v}_{\rm A})^2]} - \left(p_{\rm g} + \frac{4}{3}E \right)v_{\rm ph} \right].\nonumber\\
\end{eqnarray}
Equation~(\ref{eq:growth_rate_first_order}) is identical to the result obtained by \citetalias{BS03}
in the limit of rapid heat exchange (their equation 93).

The stability criterion is obtained by requiring that the last squared bracket in equation~(\ref{eq:growth_rate_first_order})
is positive. Ignoring angular factors, we can write
\begin{equation}
\label{e: easy_criteria}
F \gtrsim \Xi\left[\omega,{\bf k},{\bf B}\right]\, \left(p_{\rm g} + \frac{4}{3}E \right)\,v_{\rm ph},
\end{equation}
where
\begin{equation}
\label{eq:Xi_app_definition}
\Xi\left[\omega,{\bf k},{\bf B}\right]\equiv\frac{v_{\rm ph}^2-(\hat k\cdot \mathbf{v}_{\rm A})^2}{v^2_{\rm A}}
\end{equation}
measures the ratio of compressional energy to kinetic energy due to motion
along ${\bf B}$.

\begin{figure}
\includegraphics*[width=0.5\textwidth]{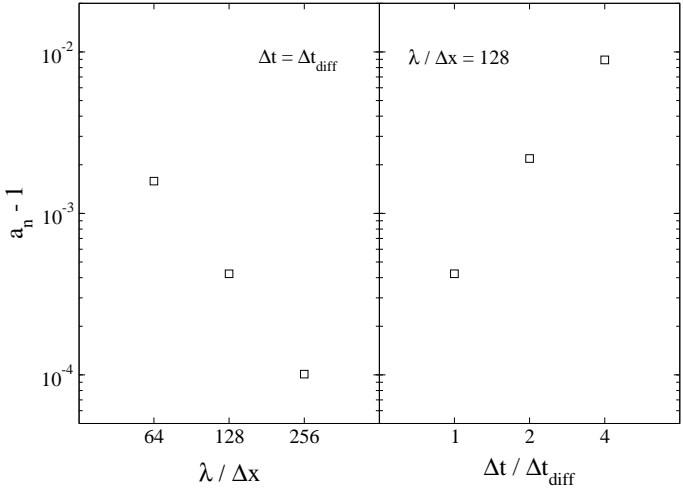}
\caption{Deviation from unity of the normalized projection $a_{\rm n}$ of the numerical solution
to the diffusion equation (eq.~[\ref{eq:diffusion_projection}]), evaluated at $t=t_0$. The
left panel shows results for different spatial resolutions, using a time step equal to
the diffusion time at the grid scale. The right panel shows results with 128 cells per side of the square
domain, using a minimum time step equal or larger than the diffusion time step.}
\label{f:diffusion_convergence}
\vspace{0.1in}
\end{figure}

\begin{figure*}
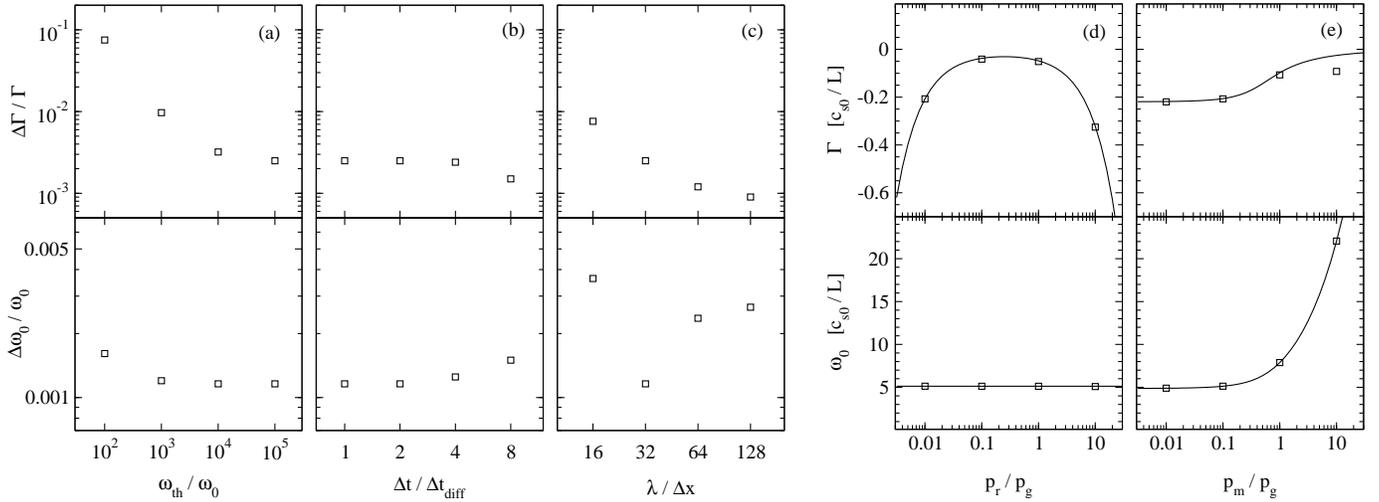

\includegraphics*[width=0.6017\textwidth]{f11a.eps}
\includegraphics*[width=0.3983\textwidth]{f11b.eps}
\caption{Properties of rapidly-diffusing, thermally-locked radiation-magnetosonic modes on a uniform background,
as a function of system parameters. Panels (a)-(c) show the fractional difference
between the fitted and analytic eigenfrequencies (upper and lower panels for damping rate $\Gamma$
and oscillation frequency $\omega_0$, respectively) obtained when varying the degree of thermal
locking and numeric parameters. The baseline model 
has $p_{\rm r}/p_{\rm g}=10^{-2}$, no magnetic field ($v_A = 0$), $\omega_{\rm diff}/\omega_0=100$,
$\omega_{\rm th}/\omega_0=10^5$, 
$\lambda/\Delta x=32$, and timestep equal to the diffusion time at the grid 
($\Delta t=\Delta t_{\rm diff}$). Panels (d)-(e) show eigenfrequencies as a function
of the relative strength of gas, radiation, and magnetic pressure. The baseline model
is the same as for panels (a)-(c) but with a non-zero magnetic field
($p_{\rm m}/p_{\rm g}=0.1$ in panel d).
Curves show analytic values (eqns.~\ref{eq:damping_uniform} and \ref{eq:disprel_magnetosonic})
while symbols show results from fits to simulations.
}
\label{f:rmhd_waves_uniform}
\end{figure*}

\section{Tests of Zeus-MP}
\label{s:code_tests}

The magnetohydrodynamic capabilities of Zeus-MP have been extensively tested \citep{hayes2006}.
However, the radiation module has only been tested when coupled to the hydrodynamics solver.
Here we describe additional tests that quantify the accuracy with which
equations~(\ref{eq:mass_conservation})-(\ref{eq:induction_equation}) are solved.

\subsection{Diffusion Solver}

To test the removal of the default flux limiter in Zeus-MP,
we compare the analytic and numeric evolution of a sinusoidal thermal wave.
In this test all the accelerations, magnetic fields,
and absorption opacities are suppressed in the code. If the density
and gas pressure are set initially
to uniform values, and the velocities vanish, the only time-dependent
equation left is the diffusion of radiation energy density:
\begin{equation}
\frac{\partial E}{\partial t} -\nabla \cdot \left(D\nabla E\right) = 0,
\end{equation}
where $D = c/(3\kappa_{\rm F}\rho)$ is the diffusion coefficient.

Following \citet{turner2001}, we employ a periodic square box of length
$L$ and a constant diffusion coefficient. A sinusoidal thermal wave with
wavelength equal to the side of the box  should evolve according to
\begin{equation}
\label{eq:ewave_analytic}
E_a(x,y,t) = E_0 + E_1\,e^{-t/t_0}\,\sin\left(2\pi\frac{x}{L}\right)\sin\left(2\pi\frac{y}{L}\right),
\end{equation}
where $E_0$ and $E_1$ are free parameters, the subscript $a$ denotes analytic solution,
and $t_0 = L^2/(8\pi^2\,D)$ is the e-folding time of the solution. At $t=0$,
we initialize the radiation energy density according to equation~(\ref{eq:ewave_analytic}).

To diagnose the agreement between the numeric and analytic solution,
the numerical radiation energy density $E_{\rm n}$ at $t=t_0$ is projected onto the appropriate
Fourier mode and normalized
\begin{equation}
\label{eq:diffusion_projection}
a_{\rm n} = \frac{4}{L^2}\frac{e^{1}}{E_1}\int_0^L\int_0^L\, E_{\rm n}(x,y,t_0)\,
      \sin\left(2\pi\frac{x}{L}\right)\sin\left(2\pi\frac{y}{L}\right) \totd x \totd y.
\end{equation}
The quantity $a_{\rm n}-1$ is a measure of the global deviation of the numerical
solution from the analytic expectation.

A number of tests were performed using different spatial resolutions
and time steps, with an initial amplitude $E_1/E_0 = 0.5$. 
Results are shown in Figure~\ref{f:diffusion_convergence}. Agreement is
better than $1\%$ for the chosen set of parameters.

\subsection{Thermally-Locked Radiation-MHD Waves on a Uniform Background}

Here we test whether the solution of equations~(\ref{eq:mass_conservation})-(\ref{eq:induction_equation})
by Zeus-MP correctly attains the limit of rapid diffusion and rapid heat exchange
(eqns.~[\ref{eq:rapid_diffusion_condition}] and [\ref{eq:thermal_locking_condition}])
in the presence of magnetic field fluctuations.
To this end, we follow the evolution of an individual fast magnetosonic eigenmode
in a uniform square domain of size $L$ with periodic boundary conditions (see also
\citealt{davis2012} and \citealt{jiang2012} for similar tests). 

In the absence of stratification, radiation diffusion causes damping, with
an asymptotic rate (eq.~[\ref{eq:growth_rate_first_order}])
\begin{equation}
\label{eq:damping_uniform}
\Gamma_{\rm uni} = -\frac{1}{2}\tau_{\rm box}\left( \frac{c_i}{c}\right)
\left[\frac{v_{\rm ph}^2 -(\hat k\cdot \mathbf{v}_{\rm A})^2}
{2v_{\rm ph}^2 - c_{\rm i}^2 - v_{\rm A}^2}\right]\left(\frac{p_{\rm g}}{4p_{\rm r}} + 1 \right)
\left(1 + \frac{4p_{\rm r}}{p_{\rm g}} \right)\,\frac{c_i}{L},
\end{equation}
where $\tau_{\rm box}=\kappa_{\rm F}\rho_0 L$ is the optical depth in the box.
The rapid diffusion and heat exchange conditions ensure that the 
real part of the frequency $\omega_0$ is that of an isothermal magnetosonic
mode in the absence of radiation (eq.~[\ref{eq:disprel_magnetosonic}]).

The parameters of the problem are a wavelength equal to the domain size, 
wave-vector along the $z$ axis, and a magnetic field at 45 degrees to this axis.
The fiducial model has a ratio of pressures
$p_{\rm r}/p_g = 10^{-2}$, $p_m/p_g = 0.1$, an optical depth such that
$\omega_{\rm diff}/\omega_0 = 100$, a ratio of absorption to
flux-mean opacity such that $\omega_{\rm th}/\omega_0 = 10^{5}$, a
resolution of 32 cells per wavelength in both directions, and a time step equal to the diffusion
time at the grid scale. All other parameters have the default values described in \S\ref{s:model_parameters}.
The domain is initialized with a plane wave in density and gas pressure
with amplitude $\delta \rho/\rho_0 = \delta p/p_0 = 10^{-2}$, and
spatial dependence $\propto \cos(2\pi z/L)$. 
The initial magnetic field and velocity perturbations are obtained
from equations~(\ref{eq:induction_linear}) and (\ref{eq:momentum_cons_linear2}), respectively.

To quantify agreement with the analytic expectation, we project the
fractional deviation of the density from its equilibrium value onto
the appropriate Fourier component, obtaining a time-dependent coefficient
\begin{equation}
a_{\rm wave}(t) = \frac{2}{L^2}\int_0^L\int_0^L\left[\frac{\rho(x,z,t)}{\rho_0} -1\right]
\cos\left(\frac{2\pi z}{L} \right)\,\totd x\totd z.
\end{equation}
The above coefficient is then fitted with a cosine function of exponentially decreasing amplitude,
yielding an oscillation frequency, damping rate, and initial amplitude.

We evolve two series of runs. The first has no magnetic field, and samples different 
spatial resolutions, time steps, and absorption opacities. Results are shown in 
Figure~\ref{f:rmhd_waves_uniform}a-c, in the form of a fractional difference
between the analytic and fitted eigenfrequencies. Agreement is better
than $1\%$ in $\omega_0$ and $10\%$ in $\Gamma_{\rm uni}$, respectively,
for all chosen parameters. The second set has a non-zero magnetic field,
and samples different radiation, magnetic, and gas pressures. Results
are shown in Figure~\ref{f:rmhd_waves_uniform}d-e. Agreement between
analytic and fitted eigenfrequencies is excellent over a wide region
of parameter space. The only exception is the damping rate for very
large magnetic field, where a larger damping rate is measured.

\bibliographystyle{apj}
\bibliography{photon_bubble,apj-jour}

\end{document}